\documentclass[12pt]{iopart}
\usepackage{iopams}

\usepackage{graphicx}
\usepackage[utf8]{inputenc}
\usepackage[T1]{fontenc}
\usepackage{dcolumn}
\usepackage{bm}
\usepackage{iopams} 
\usepackage{color}
\usepackage{soul}
\usepackage{url}
\usepackage[normalem]{ulem}
\usepackage{aas_macros}

\usepackage{multirow}
\usepackage[toc,page]{appendix}

\usepackage[caption=false]{subfig}

\begin{document}

\title[Anomaly Detection in Gravitational Waves data]{Anomaly Detection in Gravitational Waves data using Convolutional AutoEncoders}

\author{Filip Morawski}
\ead{fmorawski@camk.edu.pl}
\address{Nicolaus Copernicus Astronomical Center, Polish Academy of Sciences, Bartycka 18, 00-716, Warsaw, Poland}

\author{Micha{\l} Bejger}
\address{Nicolaus Copernicus Astronomical Center, Polish Academy of Sciences, Bartycka 18, 00-716, Warsaw, Poland}

\author{Elena Cuoco}
\address{European Gravitational Observatory (EGO), I-56021 Cascina, Pisa, Italy}
\address{Scuola Normale Superiore, Piazza dei Cavalieri 7, I-56126 Pisa, Italy}
\address{INFN, Sezione di Pisa, Largo Bruno Pontecorvo, 3, I-56127 Pisa, Italy}

\author{Luigia Petre}
\address{Department of Computer Science, Faculty of Science and Engineering}
\address{\AA bo Akademi University, Tuomiokirkontori 3, 20500 Turku, Finland}

\date{\today}

\begin{abstract}
 
As of this moment, fifty gravitational waves (GW) detections have been announced, thanks to the observational efforts of the LIGO-Virgo Collaboration, working with the Advanced LIGO and the Advanced Virgo interferometers. The detection of signals is complicated by the noise-dominated nature of the data.  Conventional approaches in GW detection procedures require either precise knowledge of the GW waveform in the context of matched filtering searches or coincident analysis of data from multiple detectors. Furthermore, the analysis is prone to contamination by instrumental or environmental artifacts called glitches which either mimic astrophysical signals or reduce the overall quality of data.

In this paper, we propose an alternative generic method of studying GW data based on detecting anomalies. The anomalies we study are transient signals, different from the slow non-stationary noise of the detector. Presented in the manuscript anomalies are mostly based on the GW emitted by the mergers of binary black hole systems. However, the presented study of anomalies is not limited only to GW alone, but also includes glitches occurring in the real LIGO/Virgo dataset available at the Gravitational Waves Open Science Center.
To search for anomalies we employ deep learning algorithms, namely convolutional autoencoders, which are trained on both simulated and real detector data. We demonstrate the capabilities of our deep learning implementation in the reconstruction of injected signals. We study the influence of the GW strength, defined in terms of matched filter Signal-to-Noise Ratio, on the detection of anomalies. Moreover, we present the application of our method for the localization in time of anomalies in the studied time-series data. We validate the results of anomaly searches on real data containing confirmed gravitational wave detections; we thus prove the generalization capabilities of our method, towards detecting GWs unknown to our deep learning models during training.

\end{abstract}

\maketitle


\section{Introduction}
\label{sec:intro}

The first gravitational wave (GW) detection on September 14, 2015 \cite{GW150914} inaugurated a new era in astrophysics. The joint observational effort of the LIGO and Virgo Collaborations, working with the Advanced LIGO \cite{ligosc2015} and the Advanced Virgo \cite{Acernese_2014} interferometers in a global network has provided, until the suspension of the LIGO-Virgo observing run O3 in Spring 2020, fifty GW candidate signal detections \cite{ligoCIT,virgo,GraceDB,gwtc-2}. Such an impressive number of statistically-significant signal candidates allows for the verification of many theoretical models, describing various sources of GW radiation, like binary systems of black holes (BH) and neutron stars (NS), as well as the very nature of gravity \cite{Abbott_2019}. As a result of continuous work to improve their sensitivity, the Advanced LIGO and the Advanced Virgo interferometers will soon probe a much larger volume of space and expand the capability of discovering new GW sources. According to theoretical predictions, tens of BH mergers and a few NS mergers will be soon routinely registered every year \cite{abbott_2016}. Such a large number of events will deliver extraordinary information about the nature of those objects, phenomena, as well as the space-time itself.

What can be observed through the GW window strongly depends on the fidelity of the data analysis methods. The GW data is always noise dominated: astrophysical signals are buried deep into the detectors’ noise, caused by various sources. Some of them are associated with the environment \cite{Abbott_2019}, e.g., the seismic and environmental activity; others originate from the detector itself \cite{abbott_2016}, such as the thermal fluctuations of the mirrors and the laser beam photon shot noise \cite{kumar}  - time-dependent fluctuations in the laser interacting with the mirrors of the interferometer. In order to reveal the hidden GWs, the traditional approach is based on matched filtering algorithms \cite{owen_sathya}. The examples of existing pipelines utilising a matched filtering approach are PyCBC \cite{pyCBC_pipe} and GstLAL \cite{gstlal}. Assuming the model gravitational waveform is known, which is not always the case for all astrophysical sources, the algorithm scans the data to match an optimal template using a template bank. Matched filtering is computationally expensive and in case of large template banks requires a lot of computational resources. As an optimal method of filtering in stationary Gaussian noise, it is prone to contamination by non-stationary instrumental artifacts, the so-called glitches, which either mimic GW signals or reduce the quality of the collected data. Existing alternatives to matched filtering methods are based on unmodeled searches of GW burst signals. Such GWs are of short duration with an unknown or partially modelled waveform morphology related to the complicated or unknown astrophysics. These signals are searched by measuring an excessive power in the time-frequency domain that occurs coherently between multiple detectors. An example of GW burst pipeline is coherent Wave Burst (cWB) \cite{2016PhRvD..93d2004K, 0264-9381-25-11-114029}. The sensitivity of this method is affected by short duration glitches that may occur coincident in time between multiple detectors. Versatile and rapid pipelines are needed to deal with non-stationary noise (and instrumental glitches) and to perform preliminary analysis of large amounts of data from multiple detectors. 

Deep learning (DL)~\cite{Goodfellow-et-al-2016} fits in that role perfectly. DL has commenced a new era of machine learning (ML), a field of computer science based on specially-designed algorithms that can generalize (`learn') from examples in order to solve problems and make predictions, without the need of being explicitly programmed \cite{samuel_ml}. DL algorithms, based on the concept of neural networks - models of neuron connectivity in the brain - are able to analyse different representations of data with varied dimensionality, like images (spectrograms, Q-transforms, Wavelet-transform) or time series. Moreover, they can quickly process large amounts of data - a requirement for the real-time (low latency) analysis at the GW interferometers.

The purpose of this work is to propose a generic, model-independent data analysis method that searches for anomalies in signals recorded by the GW detectors, based on performant DL models. The innovation of the proposed method is that our DL model `learns' the features of the noise and detects if anomalies occur in the detected signal. We define an anomaly as an extraordinary, sparse transient data feature, outstanding with respect to the `normal' background noise of the detector.
The anomaly signal could therefore either be represented as a GW or an instrumental glitch. In particular, the GWs studied in the presented work are based on the signals emitted by the binary BH systems (BBH).

The term `anomaly' is a well defined concept in statistics and data analysis (hence also in machine learning), sometimes also referred to as `outlier'. The DL methods we apply are fitting perfectly in their detection and analysis role here, being especially suited to detect, compress, and reconstruct non-linearities in the input signal data.
The analysis assessing the capabilities of the method is performed with the BBH signals, injected (added) to both simulated and real detector data. Once trained and tested on injections, the DL models are validated on the confirmed GW detections - real astrophysical signals registered by the detectors. Our results provide a proof-of-concept for the advantages of using DL methods in the GW data analysis, namely the processing speed and ability to capture complicated non-linear relationships in the data. These features enable our method to be potentially used as an Event-Trigger-Generator (ETG). In this context, the advantage of our method is the computational speed - ML algorithms once trained are extremely efficient in the processing of data, in particular when used on computers equipped with a Graphics Processing Units (GPU). By searching for anomalies in the gravitational waves' data via ML algorithms, our method could support the currently existing ETG such as Omicron \cite{omicron} and  Q-transform based Omega \cite{omega_qtransform}.

In the GW astronomy, while DL is being actively researched, it is still quite a novel method. Therefore this research fits very well in the early adoption scheme of the modern state-of-the-art development in the field, a knowledge of which will become indispensable in the nearest future.  In the following we mention a few interesting test cases. George et al. \cite{GEORGE201864} developed the Deep Filtering method for signal processing, based on a system of two deep convolutional neural networks, designed to detect and estimate parameters of compact binary coalescence signal in highly noisy time-series data streams. The same authors have been involved in a group working on denoising gravitational waves with autoencoders \cite{shen}. Dreissigacker et al. \cite{PhysRevD.100.044009} have been using DL as a search method for the continuous GW emitted by spinning neutron stars. Similar work has been conducted by Morawski et al. \cite{morawski_2020} on the application of convolutional neural networks for classifying continuous GW signal candidates. Furthermore, DL has been used by Beheshtipour and Papa \cite{dl_cont_gw} for clustering continuous gravitational wave candidates. DL has also been successfully used in the classification of glitches by Razzano and Cuoco \cite{Razzano_2018}. Finally DL has been used in searches for GW emitted by core-collapse supernova explosions by Iess et al. \cite{Iess_2020}.

Within the GW astrophysics, application of DL in anomaly detection is a fairly new concept (see however a recent application in \cite{CORIZZO2020113378}). In different fields of astrophysics the ML anomaly detection has already been successfully implemented \cite{2019MNRAS.484..834G,daddona2020anomaly} (see \cite{2019arXiv190407248B} for a recent general overview of ML in astronomy). Outside astronomy, anomaly detection has also been  proposed in search for signatures of new physics in the Large Hadron Collider data \cite{PhysRevD.101.075021}, where it was demonstrated to discover a specific class of highly-energetic particle jets, without the prior knowledge of their specific features. 

The outline of this work is as follows. In Section~\ref{sec:dl} we briefly discuss the DL architecture applied in our work and in Section~\ref{sec:meth} we describe our data generation procedures. In Section~\ref{sec:results} we explain the results of the anomaly detection studies performed on both simulated and real data. Finally, in Section~\ref{sec:summary} we further discuss our results and draw some conclusions.

\section{Deep Learning algorithms}
\label{sec:dl}

In this paper we employ a combination of two deep learning methods (CNNs and AEs) for distinguishing GW from noise signals. We first briefly describe these learning methods and then we explain how we applied them to our problem.

\subsection{CNN and AE}
\label{ssec:dl}

\begin{figure}[htbp]
  \centering
  \includegraphics[width=1\columnwidth]{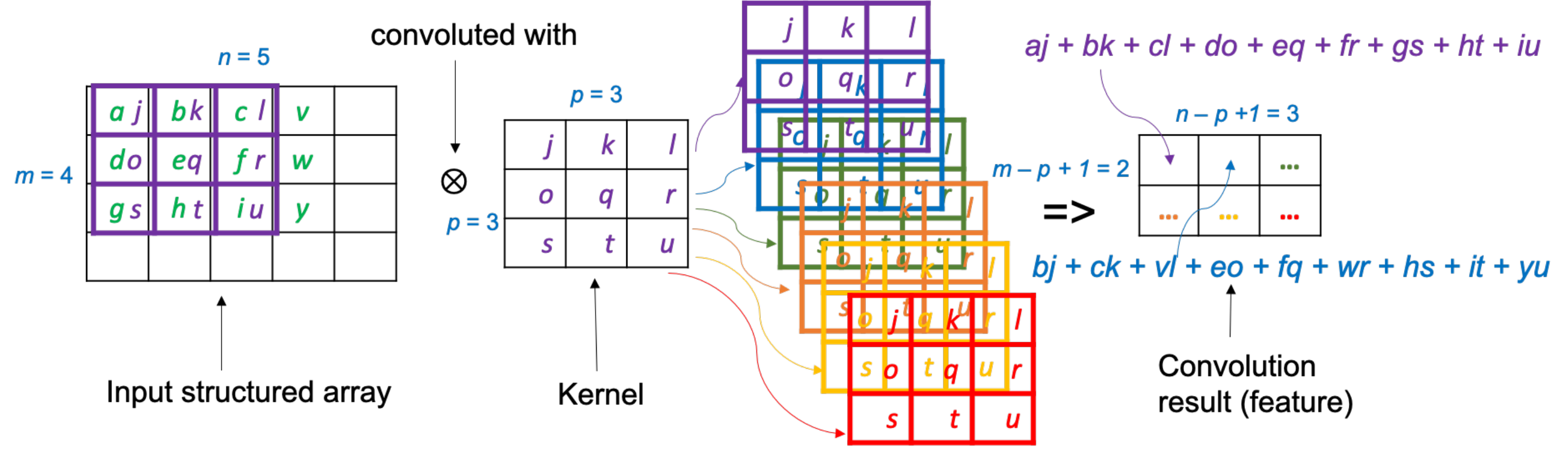}
  \caption{The 2-dimensional 4$\times$5 array on the left is seen as 6 overlapping 3$\times$3 arrays; each of these smaller arrays is inner-multiplied with the 3$\times$3 kernel, resulting in the 2$\times$3 array on the right.
}    
  \label{fig:convoluted} 
\end{figure}

A \emph{convolutional neural network} \cite{Goodfellow-et-al-2016} (CNN) is a deep, feed-forward artificial neural network (processes the information one-way, from the input to the output), designed for processing structured arrays of data, e.g., for classifying images. The core feature of a CNN is the convolution operator, that differentiates them from regular (linear) neural networks employing simple matrix multiplication. The convolution operation envisions the input structured array (say, of dimension $m\times n$) as a sequence of overlapping elements of dimension, say, $p\times p$, often with $p < min(m,n)$. The convolution operator inner-multiplies each such element with a so-called kernel (or filter), of the same dimension ($p\times p$), that `slides' over each element of the original array. Each $p\times p$ element of the input array is thus replaced by the result of the convolution operation (a number), and we thus have dimensionality reduction. This somewhat simplified convolution concept is illustrated in Figure~\ref{fig:convoluted}.

Multiple layers are typically sequentially used, each with varying numbers and types of kernels. The kernels are chosen so that the network learns specific features in one convolution (e.g., edges, corners, etc). Convolution layers are alternated with other specialised layers, all further reducing the dimensionality, until a final activation function maps the last layer to a vector whose elements correspond to the desired options (classes) for classifying the input structured array. Each class in this vector is usually a probability, so that the class with the highest value is indicated as the recognised class.

The overlapping of the input array elements together with the sliding kernel, as well as the sequencing of applying convolution step-wise simulates the structure and operation of the visual human cortex that processes incoming images in a series of layers, by identifying progressively more complex features. CNNs have been shown to work extremely well at picking up patterns in the input structured arrays, for instance in images, and thus have rapidly become the state-of-the-art in image classification and computer vision.

In our approach we use all the advantages brought by CNNs, but we further embed them into an \emph{autoencoder} (AE) architecture \cite{baldi}. An AE \cite{Goodfellow-et-al-2016} is a special type of deep artificial neural network that step-wise encodes and compresses the input and then it (re)constructs an output based only on the most compressed encoding, called hidden layer, latent representation, or bottleneck. The main AE hypothesis is that some structure exists in the input data, for instance some form of correlation between the input features; such structure can be and is learnt by an AE and consequently leveraged when forcing the input through the latent layer. (If the input features are independent of each other, then the compression and subsequent reconstruction are very difficult.) 

An ideal AE is sensitive enough to the inputs to accurately build a reconstruction and insensitive enough to the same inputs so that the model does not memorize/overfit the training data. Differently formulated, this forces the model to maintain in the latent representation only the variations in data needed for reconstruction, without holding on to redundancies within the input. There are different types of AE, but in our work we use the undercomplete AE, where the dimension of the latent representation is strictly smaller than the input dimension. In this way we take care of avoiding overfitting, since the model will not be able to copy the input to the output.

\begin{figure}[htbp]
  \centering
  \includegraphics[width=0.8\columnwidth]{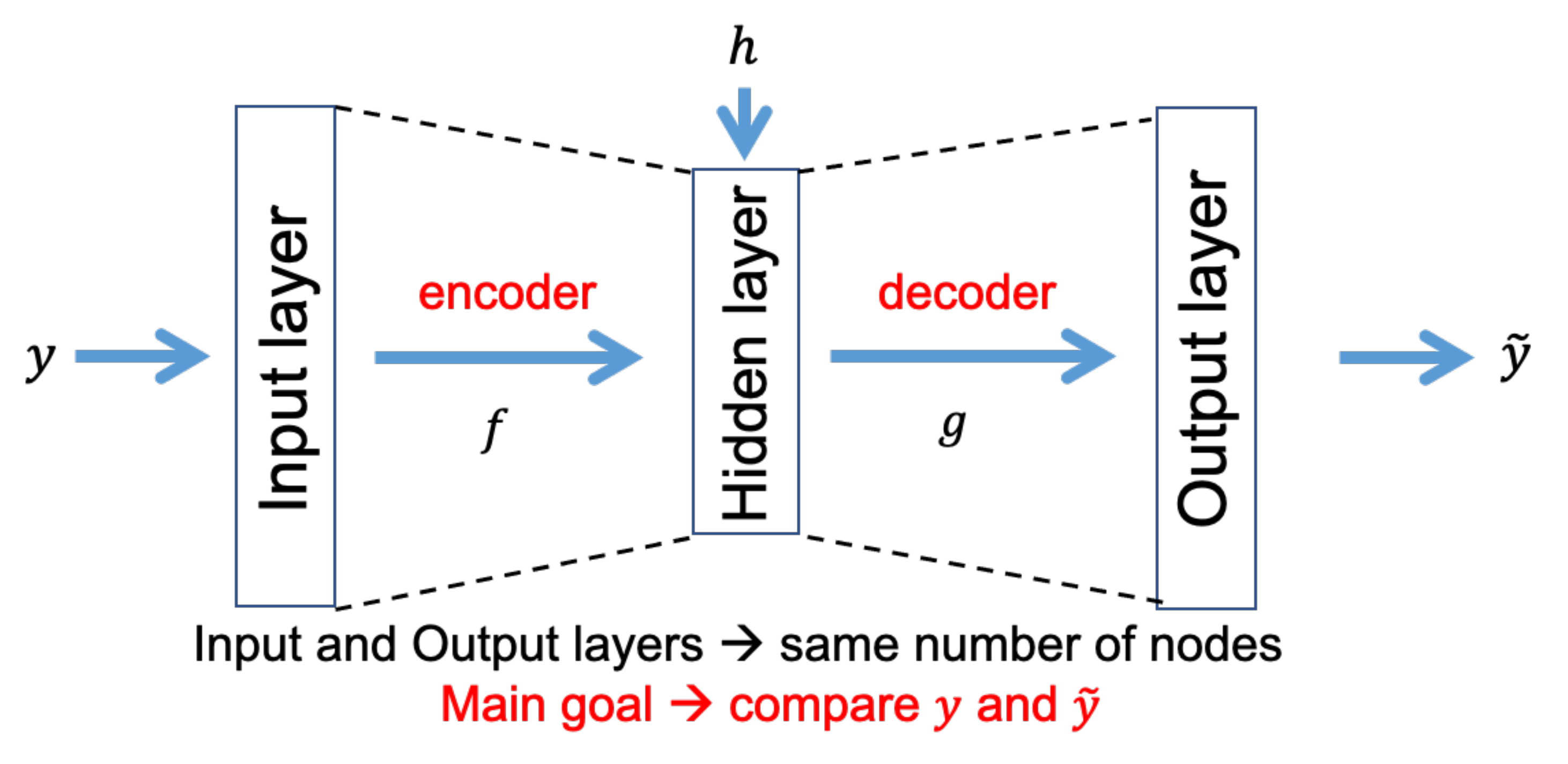}
  \caption{An AE with one hidden layer.
}    
  \label{fig:AE} 
\end{figure}

The simple AE in Figure~\ref{fig:AE} (with only one hidden layer) works by first encoding the input $y \in \mathbb{R}^d$ to the element $h \in \mathbb{R}^p$, where $p<d$. Assuming we have the encoder activation function $f:\mathbb{R}^d \rightarrow \mathbb{R}^p$, then $h=f(Wy+b)$, where $f$ can be sigmoid, {\tt ReLU}, or something else, $W$ is a weight matrix and $b$ a bias factor (initialised randomly and updated iteratively during training). The decoder activation function $g:\mathbb{R}^p \rightarrow \mathbb{R}^d$ will map $h$ to $\tilde{y}=g(\tilde{W}h+\tilde{b})$, where the activation function $g$, the weight matrix $\tilde{W}$ and the bias factor $\tilde{b}$ may be unrelated to $f$, $W$, and $b$. If only {\tt ReLU} is used for activation and we have only one hidden layer, then we have a linear AE; if we have more hidden layers, or non-linear activation(s), then the AE becomes non-linear, which is better at detecting abstract features. Thus, in general, the encoder and decoder are proper neural networks in themselves, not simply activation functions.

The training of the AE works by attempting to minimize the reconstruction loss, most often using the mean squared error (MSE) formula~\cite{Goodfellow-et-al-2016} 

\begin{equation}
\label{eq:loss} 
{\cal L}(y,\tilde{y}) = \left\Vert{y-\tilde{y}}\right\Vert^2 = \left\Vert{y-g(\tilde{W}f(Wy+b)+\tilde{b})}\right\Vert^2.
\end{equation}

\noindent 
The training of the network works by updating the parameters $W$, $b$, $\tilde{W}$, $\tilde{b}$ until ${\cal L}(y,\tilde{y})$ is sufficiently small and further training does not decrease it anymore - in that case the network has converged. There are numerous algorithms for updating the parameters, but here we use the ADAM algorithm (adaptive moment estimation)~\cite{adam2014}, as it adapts the learning rate during training and has been empirically shown superior to other methods for large datasets, large number of parameters, as well as non-stationary input. The learning rate (training `step' of updating) together with the batch size (number of training examples that use the same learning rate), the update method (also called as optimizer), the number and size of the layers and the loss function are the hyperparameters of the AE.

\begin{figure}[htbp]
  \centering
  \includegraphics[width=0.6\columnwidth]{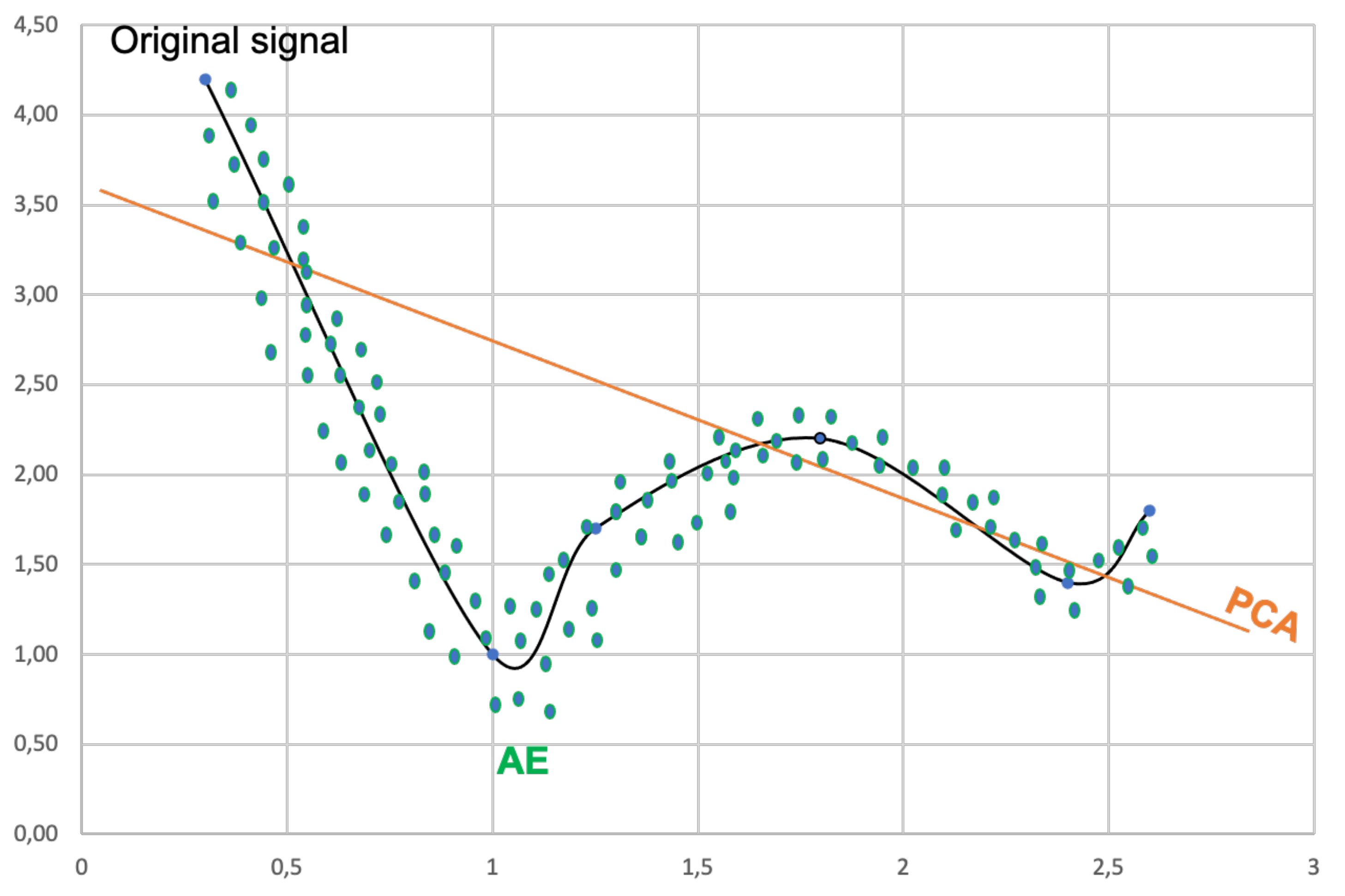}
  \caption{Example of the manifold concept. Non linear vs linear dimension reduction. 
}    
  \label{fig:manifold} 
\end{figure}

Since the AE is reducing dimensionality for encoding, it has often been used for only that - for instance, for feature learning. However, when compared to other dimensionality reduction techniques, such as Principal Component Analysis (PCA), AE appears as a powerful generalization, since it is able to learn non-linear relationships in input data. While PCA attempts to discover a lower dimensional hyperplane describing the original data, AE is capable of learning non-linear manifolds of the least possible size, as illustrated in Figure~\ref{fig:manifold}. Essentially, the AE learns a vector field for mapping input data towards lower dimensional manifolds, that describe the high density region where input data concentrates. If the manifold accurately describes input data, then the AE has effectively learnt the input data.

\subsection{Applying the CNN AE}
\label{ssec:dlApplied}

Although CNNs were designed for the analysis of 2D data (i.e., images) \cite{Goodfellow-et-al-2016}, here we apply them for a simpler, 1D implementation, as our analysed signals are time series. The CNN AE is designed to learn two things. In case of input data instances containing only detector background noise (no anomalies), the AE is trained to reconstruct the noise as closely as possible. However, in case of instances containing an anomalous signal (GW or glitch in case of real data),
the AE is trained to disregard the anomaly and reconstruct the data as if the signal was not present. By comparing the input to the output reconstructed by the AE, the anomaly present in the instance of data time series is recovered and further studied. Using the AE network is thus instrumental, since it will reconstruct only the signals it has been trained for, in our case the detector noise, and disregard anything else in its reconstruction, in our case the GW or the glitch, as noise.

\begin{figure}[htbp]
  \centering
  \includegraphics[width=0.5\columnwidth]{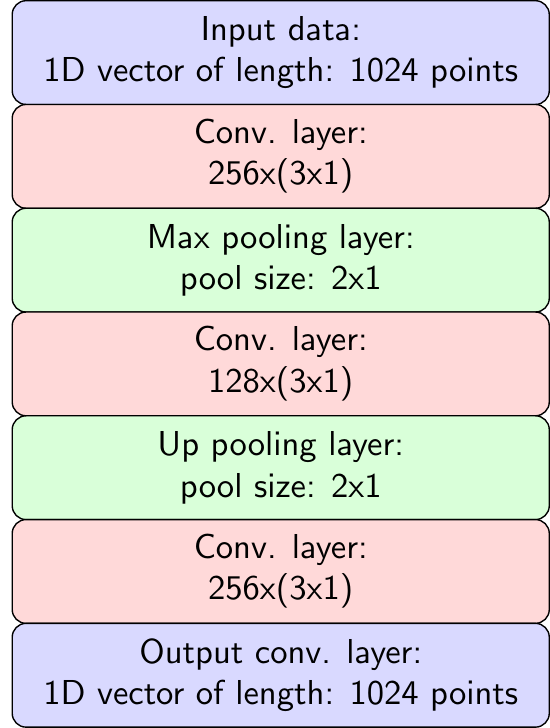}
  \caption{Diagram shows the networks’ layer structure and architecture. The size below convolutional layers correspond to the \textit{Number of filters} times \textit{Kernels size} of every layer. The second dimension of all layers is equal to unity since the initial input data were 1D time-series vectors.
}    
  \label{fig:ae_arch} 
\end{figure}

Depending on the presence of anomalies in the input data, the loss value ${\cal L}(y,\tilde{y})$ is expected to vary. The ${\cal L}(y,\tilde{y})$ computed for an `anomalous' input reaches higher values than in case of `anomaly-free' input, since the difference between $y$ (noise or noise with anomaly) and $\tilde{y}$ (only noise) is larger. The difference between these signals is proportional to the amplitude of the anomaly. As a result, the AE trained on data containing stronger anomalies is expected to converge during training towards higher ${\cal L}(y,\tilde{y})$.

The final architecture\footnote{We implement the algorithm in {\tt Python} \cite{10.5555/1593511} using the {\tt Keras/TensorFlow} library \cite{chollet2015keras,tensorflow} with the GPU support. The development was performed on the NVidia Quadro P6000 (sponsorship via the NVidia GPU seeding grant). The production runs were deployed on the Prometheus cluster (Academic Computer Centre CYFRONET AGH) equipped with Tesla K40 GPU nodes, running CUDA 10.0 \cite{cuda} and the cuDNN 7.3.0 \cite{cuDNN}.} used in the Section~\ref{sec:results} was chosen based on empirical tests on the data. We tested architectures ranging from 1 to 8 hidden layers. The final layout of the architecture is presented in Figure~\ref{fig:ae_arch}. The chosen architecture allowing to reach the minimum value for ${\cal L}(y,\tilde{y})$ for the fixed training data set was the network containing 3 hidden convolutional layers: an encoding layer, a decoding layer, and a latent representation layer in between them, with 256, 128, and 256 neurons, respectively. The kernel size was fixed for all layers to 3$\times$1. All but the final output layer use the {\tt ReLU} as the activation function, whereas the final layer reconstructing the initial signal uses a sigmoid activation function. The other hyperparameters used for training were the ADAM optimizer \cite{adam2014} with learning rate of 0.0005 and batch size of 32.

In the following section, we detail our datasets used for training, validation and testing.

\section{Training data sets and data flow}
\label{sec:meth}

\begin{figure}[htbp]
  \centering
  \includegraphics[width=\columnwidth]{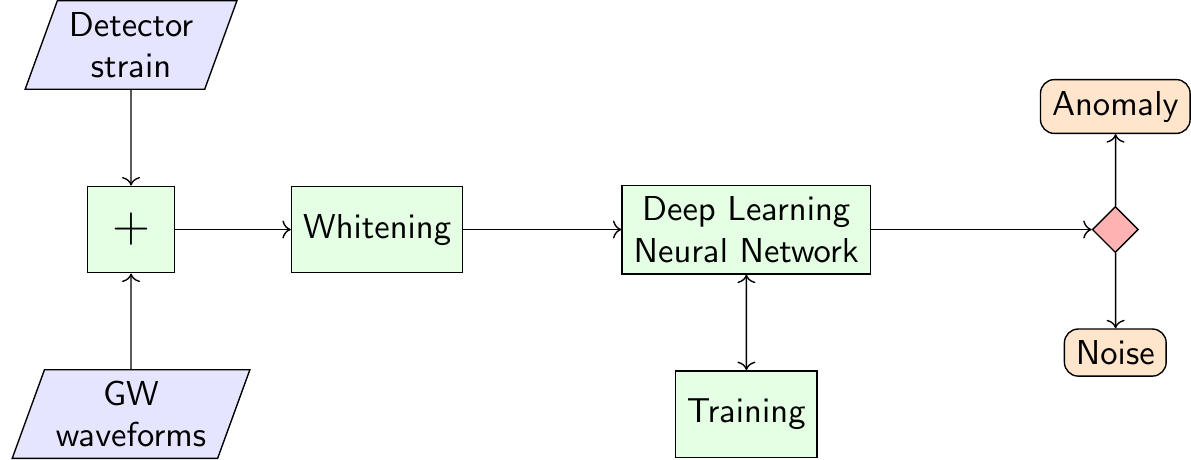}
  \caption{Data flow of the project. The first step consists of data generation. GW waveforms were injected into the detector strain, either simulated or real (e.g. public data from the GWOSC platform). 
  Since the raw time-series signal varied with frequency, the whitening procedure was applied to simplify data for further training. The training of the DL algorithms aimed to recover as many injected anomalies as possible, while limiting the false positive rate of the pipeline (noise samples incorrectly classified as anomaly).}
  \label{fig:workflow} 
\end{figure}

We prepared two kinds of training datasets: a simplified one by means of simulated detector strain time series (based on the colored normal distribution of noise), denoted DataSet 1 (DS1), and a realistic one based on the real LIGO-Virgo O2 observing run \cite{2019PhRvX...9c1040A}, publicly available at the Gravitational Waves Open Science Center (GWOSC) \cite{gwosc} and denoted DataSet 2 (DS2). In both cases we use the same general data flow presented in  Figure~\ref{fig:workflow}. 

The whitening mentioned in the workflow diagram removes the contribution of the stationary detector noise and re-weights the sensitivity at different frequencies \cite{Cuoco_2001}. As a result, the amplitude spectral density of the data becomes uniform and GW signals buried in the data are easier to search for and compare with each other. The whitening filter was re-computed separately for DS1 and DS2 as well as for every interferometer to take into account the differences in the sensitivity.
The whitening procedure used for the analysis of both studied datasets was conducted in the frequency domain following pyCBC Python library modules \cite{pyCBC}.

\begin{figure}[htbp]
  \includegraphics[width=\columnwidth]{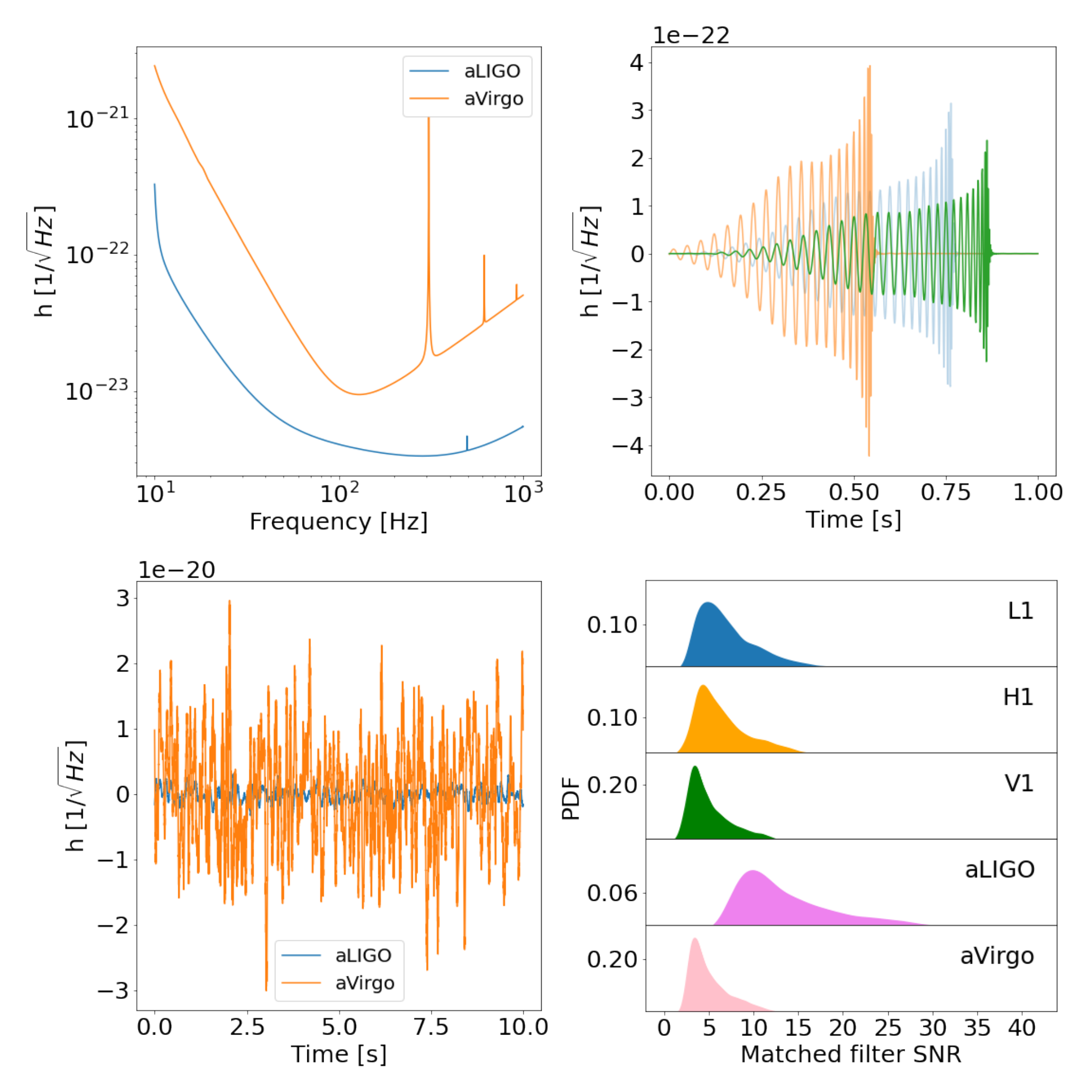}
  \caption{\textit{Top left}: Designed sensitivity of aVirgo and aLIGO interferometers. The detectors were expected to reach this level of sensitivity over broad range of frequencies after all planned upgrades. \textit{Top right}: Examples of generated BBH GW waveforms injected into the strain data as anomalies. The GW were generated for the following parameters: blue signal - $m1=29$ $M\odot$, $m2=24$ $M\odot$, $distance=440$ $Mpc$; orange signal - $m1=33$ $M\odot$, $m2=27$ $M\odot$, $distance=380$ $Mpc$; green signal - $m1=28$ $M\odot$, $m2=23$ $M\odot$, $distance=600$ $Mpc$. \textit{Bottom left}: Examples of the simulated data using the above sensitivity curves. \textit{Bottom right}: Distributions of matched filter SNR of simulated GW injected into the real data for detectors: LIGO Hanford (blue), LIGO Livingston (orange) and Virgo (green) as well simulated data for: aLIGO (violet) and aVirgo (pink). 
}    
  \label{fig:overview} 
\end{figure}

To simulate an astrophysical GW signal emitted by a BBH we used the IMRPhenomv4 waveform model \cite{PhysRevLett.113.151101}, which includes the binary inspiral, merger of the components and the final BH ringdown. The component BH masses $m_1$ and $m_2$ of the waveform model were chosen to be compatible with the first detected GW150914 \cite{GW150914}. We selected $m_1$, $m_2$ based on the Initial Mass Function (IMF)
in ranges associated with uncertainties of mass estimation for GW150914: $m_1: 32.5-40.3$ $M_\odot$, $m_2: 26.2-33.6$ $M_\odot$. For the index of the IMF power law, we chose the value of $\alpha=-2.35$ \cite{Salpeter}. The luminosity distances were chosen uniformly from the range between $200$ to $800$ $Mpc$, to cover a realistic range of the matched filter Signal-to-Noise Ratio (SNR) - from 4 to 40 varying for different interferometers as shown in the bottom right plot in Figure \ref{fig:overview}. The position in the sky was chosen to be optimal for every detector in a given moment of time. The examples of a few simulated GW signals are presented on the top right plot in Figure~\ref{fig:overview}.  

The DS1 dataset was created for the two assumed sensitivity curves of the GW detectors. Each curve described the level of the detector sensitivity with respect to the frequency in such a way that the generated strain was mimicking the realistic time series output. In our analysis we used the designed sensitivity for the advanced Virgo (aVirgo) from O3 run (version without squeezing)  \cite{VO3, VO3sensitivity} and the advanced LIGO (aLIGO) \cite{aLIGO} interferometers.
`Designed' means that the interferometers were expected to reach this level of sensitivity after all the planned upgrades.
Band-pass filtering was then applied to the generated noise to remove high frequency (above 1 kHz) and low frequency (below 30 Hz, corresponding to the seismic noise) components from the data, as current interferometers are not sensitive enough to detect GWs outside that frequency range. The data was then resampled from 4096 Hz to 1024 Hz. An example of the output time series from the DS1 is shown on the bottom left plot in Figure~\ref{fig:overview}. Prepared in advance GW signals were injected into the generated strain and subjected to the procedure of whitening.

The DS2 (realistic) dataset was created based on the publicly available LIGO-Virgo O2 observing run, using the data stored at the GWOSC platform \cite{gwosc} for three interferometers: LIGO Hanford, LIGO Livingston and Virgo described in the text with the respective abbreviations $H1$, $L1$ and $V1$. For each detector, we chose six hours of data to train the DL models. For $L1$, we took segments between 1187270656 and 1187295232 (in GPS time units); for $H1$, between 1174958080 and 1174982656; whereas for $V1$, between 1187672064 and 1187696640. We injected into the strain the same GW signals as for the simulated data. 
As a result, we obtained (for the same set of injections) three different distributions of matched filter SNR, since every detector had a different sensitivity. The distributions are illustrated in the bottom right plot in Figure~\ref{fig:overview}. For the comparison, we included additional SNR distributions of the simulated datasets. The obtained real data strain with injected anomalies was then subjected to the procedure of whitening and resampled from 4096 Hz to 1024 Hz.

In total, we generated five datasets (two simulated for aVirgo and aLIGO, and three based on real LIGO Livingston, LIGO Hanford and Virgo O2 data) which were further split into one second segments and divided for the training, validation and testing datasets (65\%, 10\% and 25\% respectively). An additional test set, containing confirmed GW detections, has been created by using one hour of data around GPS time for each confirmed GW, whitened and resampled as described above. As confirmed GWs, we chose three BBH detections with the highest network SNR from O2 run: GW150914, GW170608 and GW170814 \cite{gwtc1}.

\section{Results}
\label{sec:results}

The results presented below are split into subsections. The first presents the results of the anomaly searches on the simulated dataset containing injected GW. The second subsection covers searches of anomalies in the real data of Virgo and LIGO interferometers. The last subsection presents capabilities of anomaly searches on the real data containing confirmed detections of GW.

\subsection{Anomaly searches on simulated data}
\label{ssec:res_sim}

\begin{figure}[htbp]
  \includegraphics[width=\columnwidth]{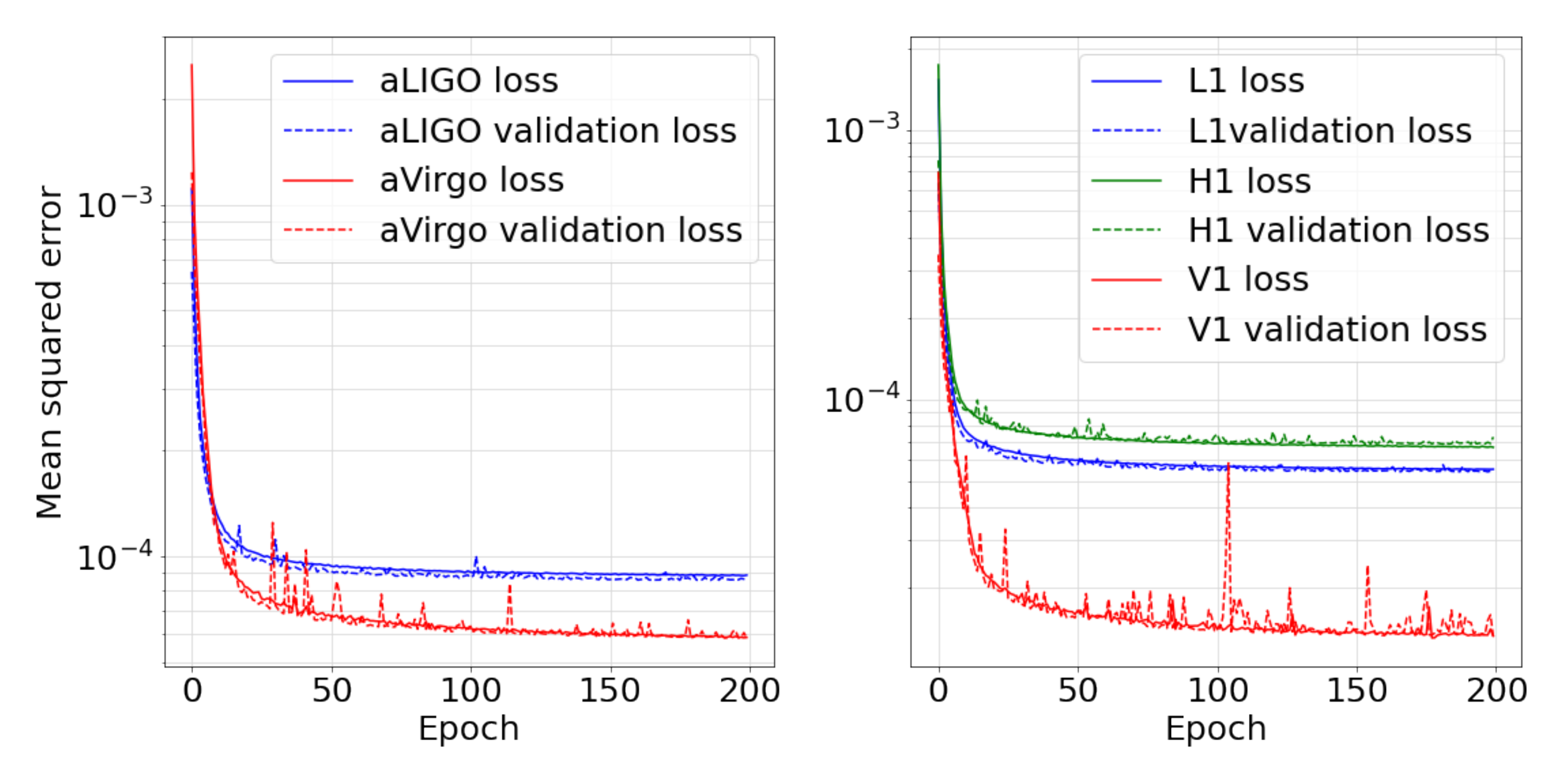}
  \caption{The learning history of the AE trained on the whitened, simulated aVirgo and aLIGO datasets (\textit{left plot}) as well as on the whitened, real data for $L1$, $H1$ and $V1$ datasets (\textit{right plot}).
  The convergence was achieved after 100 epochs. We trained for 200 epochs, to investigate the onset of overfitting: it did not appear.
}    
  \label{fig:sim_lh} 
\end{figure}

The CNN-AE described in Section \ref{sec:dl} was first trained on the whitened, simulated data containing GWs. The convergence of the model was achieved after 100 epochs, during which the MSE loss function reached a value of around $6 \cdot 10^{-5}$ for the aVirgo data and $10^{-4}$ for the aLIGO data. We extended the training for another 100 epochs to investigate the onset of overfitting. However, the overfitting didn't appear and MSE fluctuated around the values mentioned above. The learning history of the AE, trained on both simulated datasets, with results for both the training and the validation sets is shown on the left plot in Figure \ref{fig:sim_lh}. 
For aVirgo, the same set of gravitational waveforms covered a SNR range of smaller values than for aLIGO, as a result of the worse detector sensitivity (see the top left plot in Figure \ref{fig:overview} for comparison). This, in turn, resulted in the convergence towards a lower MSE for the aVirgo dataset, since the difference between the `anomalous' input and `anomalous-free' reconstruction were smaller than in the case of aLIGO data (see Section \ref{ssec:dlApplied} for more details).

\begin{figure}[htbp]
  \includegraphics[width=\columnwidth]{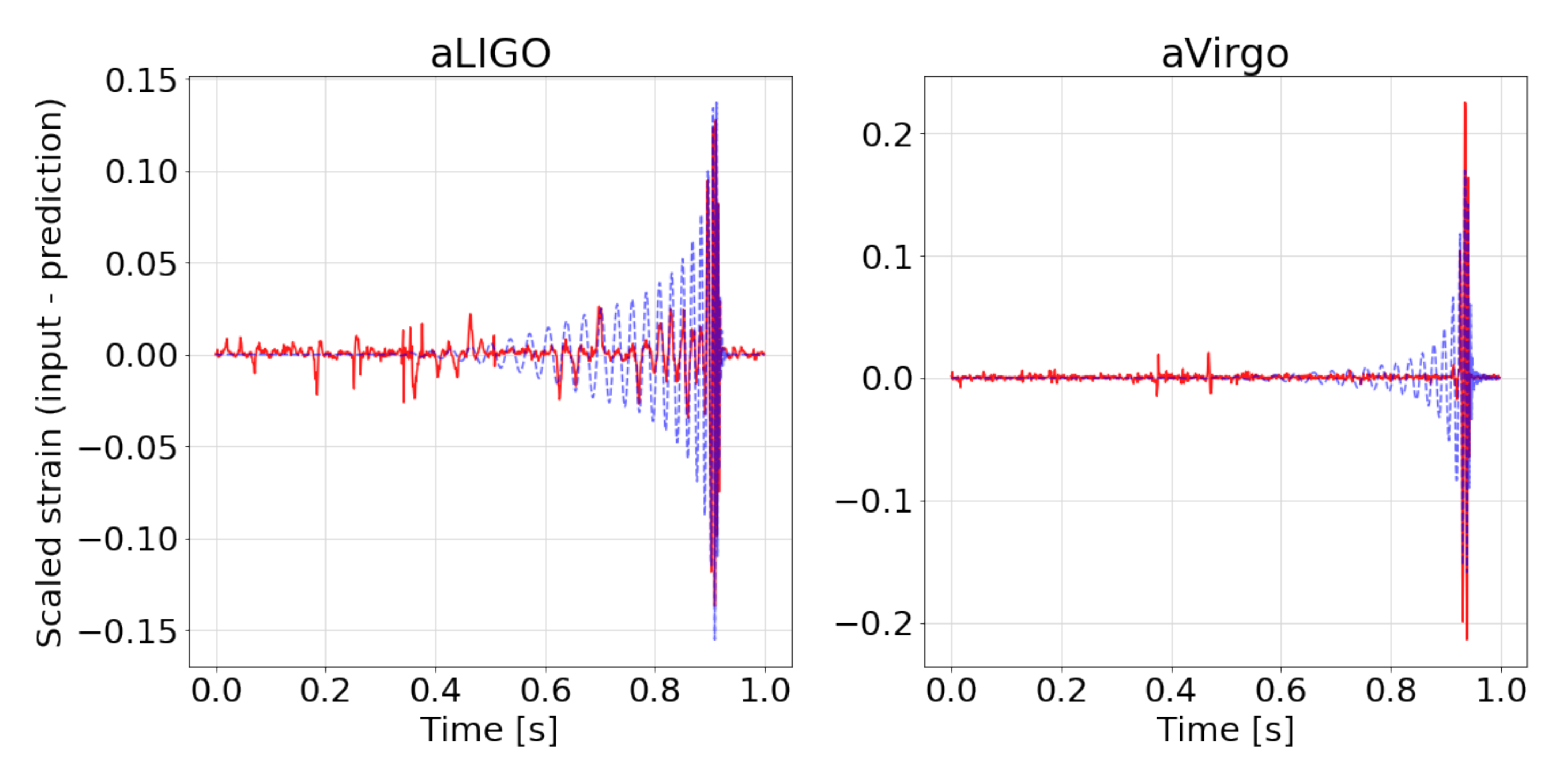}
  \caption{The examples of signal reconstruction using the CNN-AE on aLIGO (\textit{left plot}) and aVirgo data (\textit{right plot}). The plots (red curves) were generated after subtracting the input signal from the AE predictions. For comparison, we added the expected signal using blue, dashed curve (difference between the input signal and the ground-truth output signal).
}    
  \label{fig:sim_sample_rec} 
\end{figure}

Furthermore, as described in Section \ref{ssec:dlApplied}, the correctly trained CNN-AE reconstructed the pure detectors' noise, regardless of the anomalies presence in the input data. Then, by subtracting the input from the reconstructed output we aimed to recover the underlying signal.
We computed the mentioned differences between the input initial data and the output AE reconstruction. Examples of the results are presented in Figure \ref{fig:sim_sample_rec}, where blue, dashed lines in the background correspond to the expected results, whereas red lines correspond to values obtained with the AE. In the majority of cases, we were able to correctly reconstruct the GW waveforms for aLIGO data. The anomaly was significantly different from the surrounding noise, with its recovered part mostly associated with the merger part of the gravitational waveforms, together  with a small contribution from the inspiral part. In contrast, for the aVirgo dataset, the reconstruction was significantly worse, often dominated by the surrounding noise. In rare cases, as presented on the right plot in Figure \ref{fig:sim_sample_rec}, the merger part was reconstructed. In the Appendix \ref{appendix:match} we present the summary of the match between the injected and reconstructed waveforms using $<x_1 | x_2>$ metric performed in the time domain.

\begin{figure}[htbp]
  \includegraphics[width=\columnwidth]{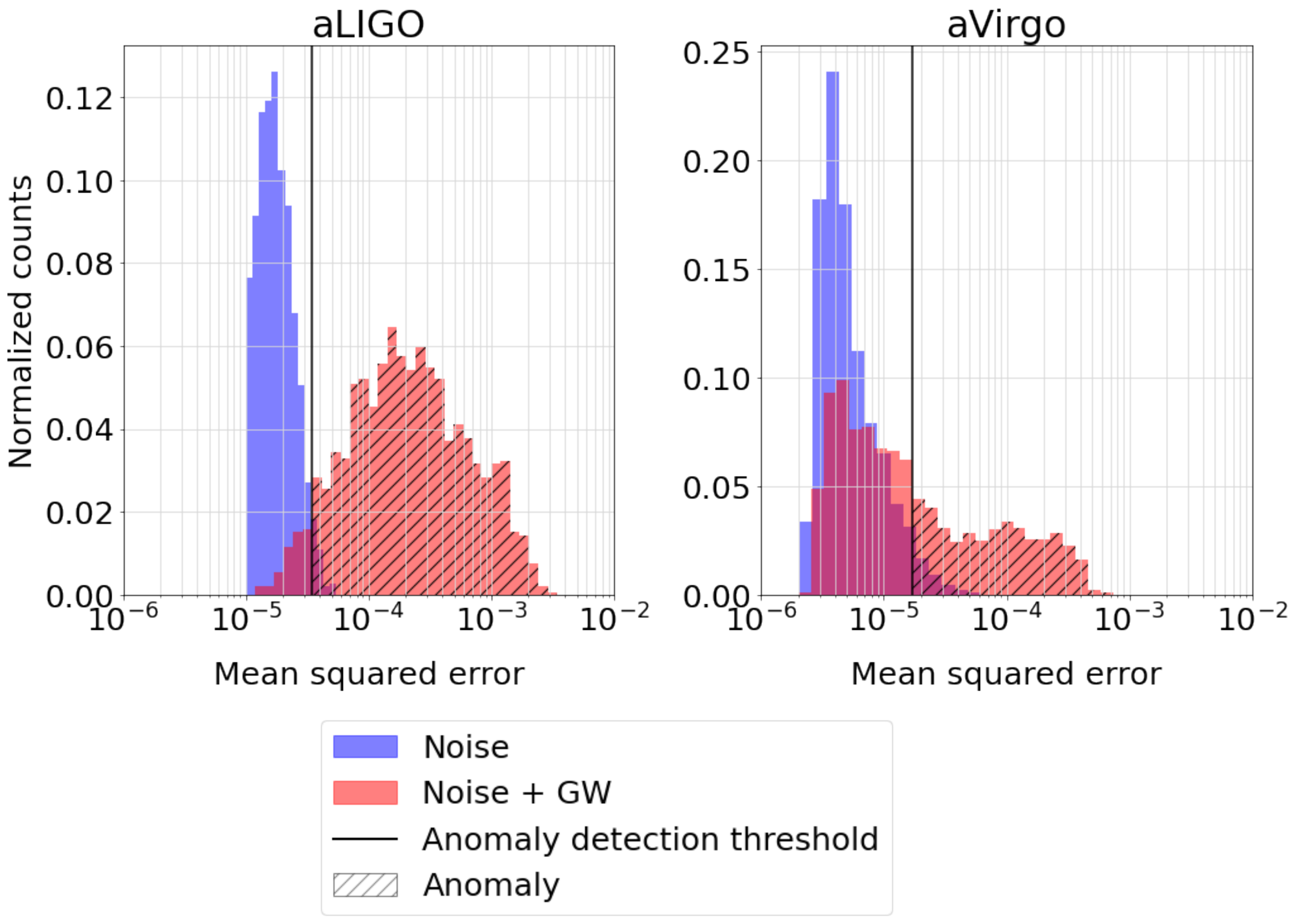}
  \caption{Distribution of MSE between the AE predictions and the input strain for two types of studied signals: noise (blue histogram) and injected GW (red histogram). In the broad range of MSE for aLIGO dataset the injected GW had significantly higher MSE than noise, allowing a definite distinction between these signal types. Additional black vertical line representing anomaly $DT$ was added to emphasize amount of detected anomalies (hatched areas).
  }    
  \label{fig:sim_hist} 
\end{figure}

Since the AE was able to correctly reconstruct the majority of anomalies buried in the data (at least for the aLIGO dataset), we aimed then to define a metric allowing automatic  anomaly detection.
We chose the MSE as such a metric, computed between the input data and the AE output as detailed in Section~\ref{ssec:dlApplied}. Figure \ref{fig:sim_hist} show histograms of MSE for the two different signal types present in the studied data: noise and injected GW. As expected, values of MSE for the noise were much smaller than for the latter case and close to zero. There was a range of MSE in which histograms of both types overlapped (denoted with the burgundy colour in Figure \ref{fig:sim_hist}). Nevertheless, the majority of noise with injected GW instances in the aLIGO dataset and almost half of these instances in the aVirgo dataset had values larger than the noise. Moreover, we added the detection threshold (defined in the following paragraph) to the histograms to stress out how many of initially injected GW were correctly detected as anomalies (hatched area in Figure \ref{fig:sim_hist}).

\begin{figure}[htbp]
  \includegraphics[width=0.5\columnwidth]{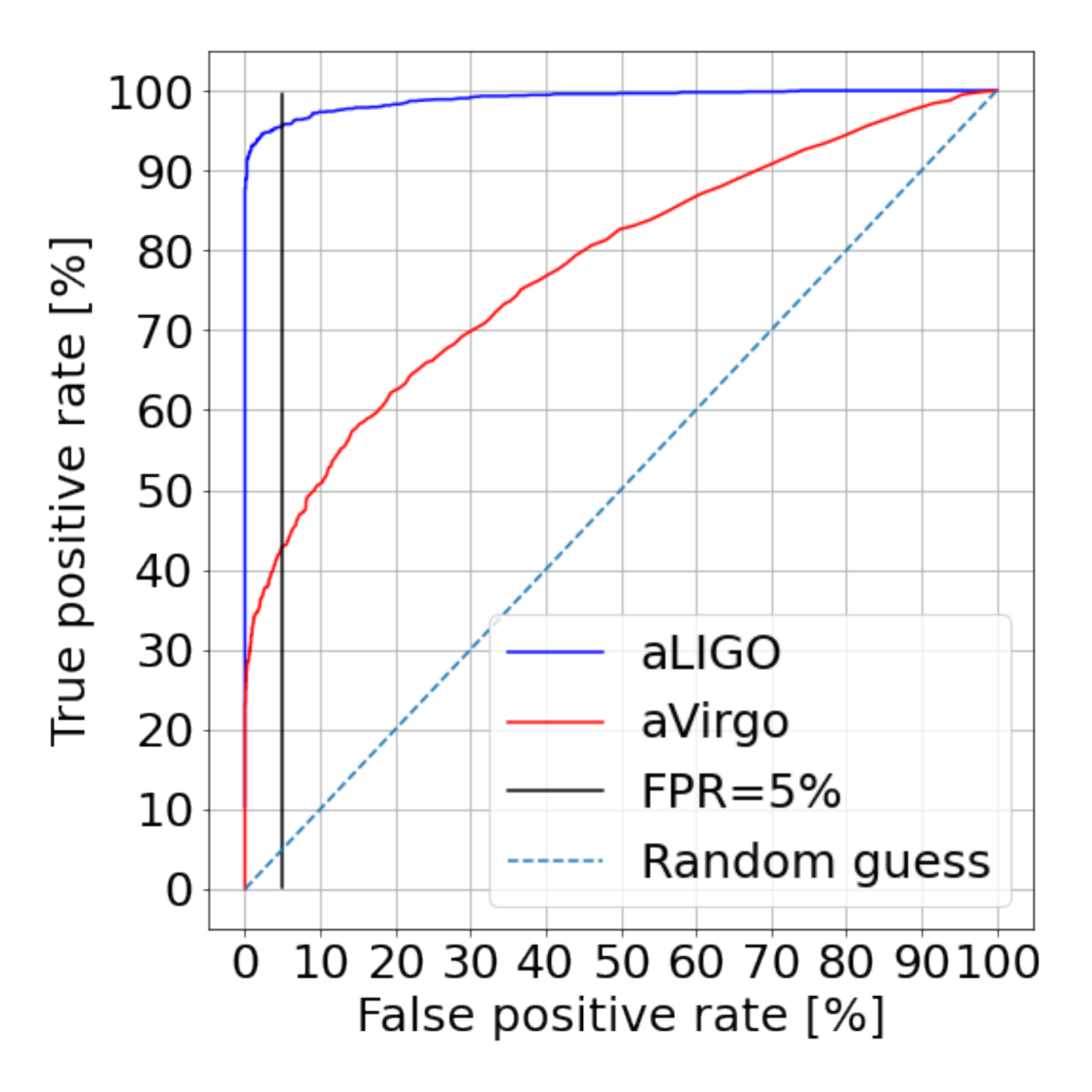}
  \hskip 10pt
  \includegraphics[width=0.5\columnwidth]{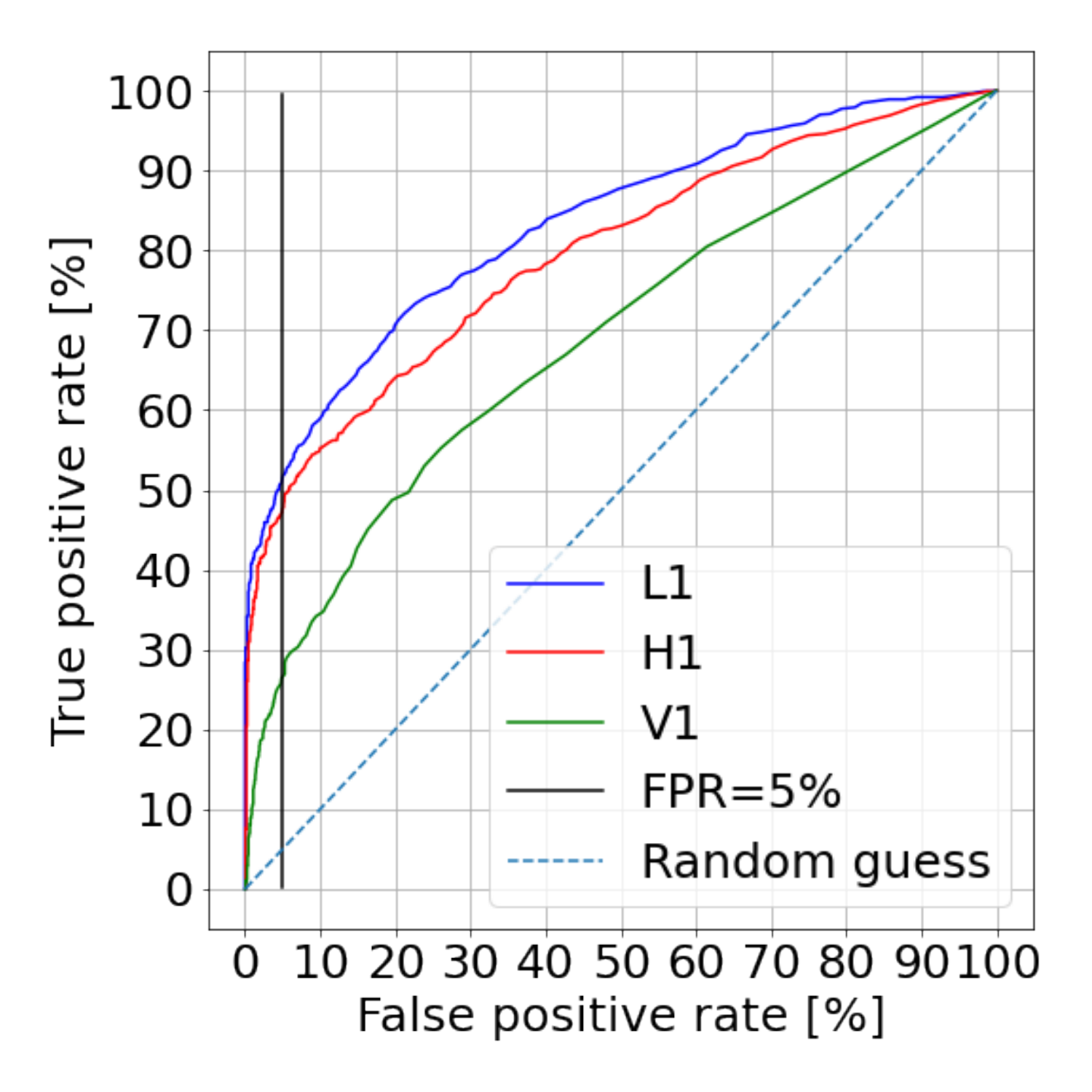}
  \caption{Receiver operating characteristic curves for simulated data (\textit{left plot}) and real data (\textit{right plot}). Black vertical line corresponds to the $FPR=5\%$ chosen as the criterion for the anomaly detection threshold.
}    
  \label{fig:roc} 
\end{figure}

We defined the threshold for the anomaly detection using the relation between the false positive rate (FPR) and MSE.
Then, by fixing FPR at a particular value, we set the detection threshold ($DT$) on the corresponding MSE. In the presented analysis, we fixed FPR at $5\%$, resulting in the following thresholds: $DT_{simV}=1.6\cdot10^{-5}$ for aVirgo and  $DT_{simL}=3.1\cdot10^{-5}$ for aLIGO. The results of the anomaly searches at FPR=5\% are shown in Table \ref{tab:sim_results} in the form of confusion matrix. Additionally, the comparison in the anomaly detection efficiency for both interferometers is shown on the left plot in the Figure \ref{fig:roc} in the form of Receiver Operating Characteristic (ROC) curves. Over all ranges of FPR, the aLIGO detector achieved a significantly higher detection efficiency, or True Positive Rate (TPR).

Presented in Tab. \ref{tab:sim_results} values quantitatively represent the results from both panels in Fig. \ref{fig:sim_hist}. Anomaly row relates to the hatched area from those panels. In the case of aLIGO dataset 96 \% of detected anomalies are correctly related to the injected GW with minor contamination of noise samples. In case of aVirgo 59 \% of samples characterized by low SNR (around 10 and less) samples did not exceed the $DT$ and contributed to the non-anomalous group. Details presenting the relation between the SNR of the injected GW and MSE of the reconstructed waveform can be found in the Appendix \ref{appendix:snr_mse}.

\begin{table*}[t]
  \centering 
  \begin{tabular}{| c || c | c || c | c |} 
    \hline
    \multirow{2}{*}{} & \multicolumn{2}{c||}{aLIGO} & \multicolumn{2}{c|}{aVirgo} \\ \cline{2-5}
      & Injected GW & Noise
      & Injected GW & Noise
      \\ \hline\hline 
    Anomaly & 96 \% & 5 \% & 41 \% & 5 \% \\ \hline 
    Non-anomaly & 4 \% & 95 \% & 59 \% & 95 \% \\ \hline
  \end{tabular}
\vskip 1em  
\caption{Results of anomaly detection of CNN-AE at FPR=5\% for aLIGO and aVirgo dataset in the form of confusion matrix. Columns relate to the ground-truth values whereas rows to the predictions. For aLIGO dataset significant majority of detected anomalies corresponded to the data samples with injected GW. However in case of aVirgo more than a half of data samples with injected GW did not exceed the $DT_{simV}$ as a result of low SNR (see Fig. \ref{fig:overview} for comparison between aLIGO and aVirgo GW SNR distributions). }
\label{tab:sim_results}
\end{table*}

\subsection{Anomaly searches on real data}
\label{ssec:res_real}

Later on, the AEs were trained on whitened, real data from the O2 observational run, collected for three existing interferometers: $V1$, $L1$ and $H1$ with injected BBH gravitational waveforms. The right plot in Figure \ref{fig:sim_lh} presents the learning history of the AE trained on the dataset for every detector. As in the case of the simulated data, AE reached convergence after around 100 epochs. We again prolonged the training to investigate the onset of overfitting, which didn't appear. Since the difference between the `anomalous' input and the `anomalous-free' reconstruction for $V1$ was the smallest among the considered datasets (as a result of the SNR smallest range), the AE during training converged towards the lowest MSE. Adequately, a smaller range of SNRs for the $H1$ dataset with respect to the $L1$ dataset resulted in the difference between the respective values of MSE (for $L1$ being higher, and for $H1$ being lower).

\begin{figure}[htbp]
  \includegraphics[width=\columnwidth]{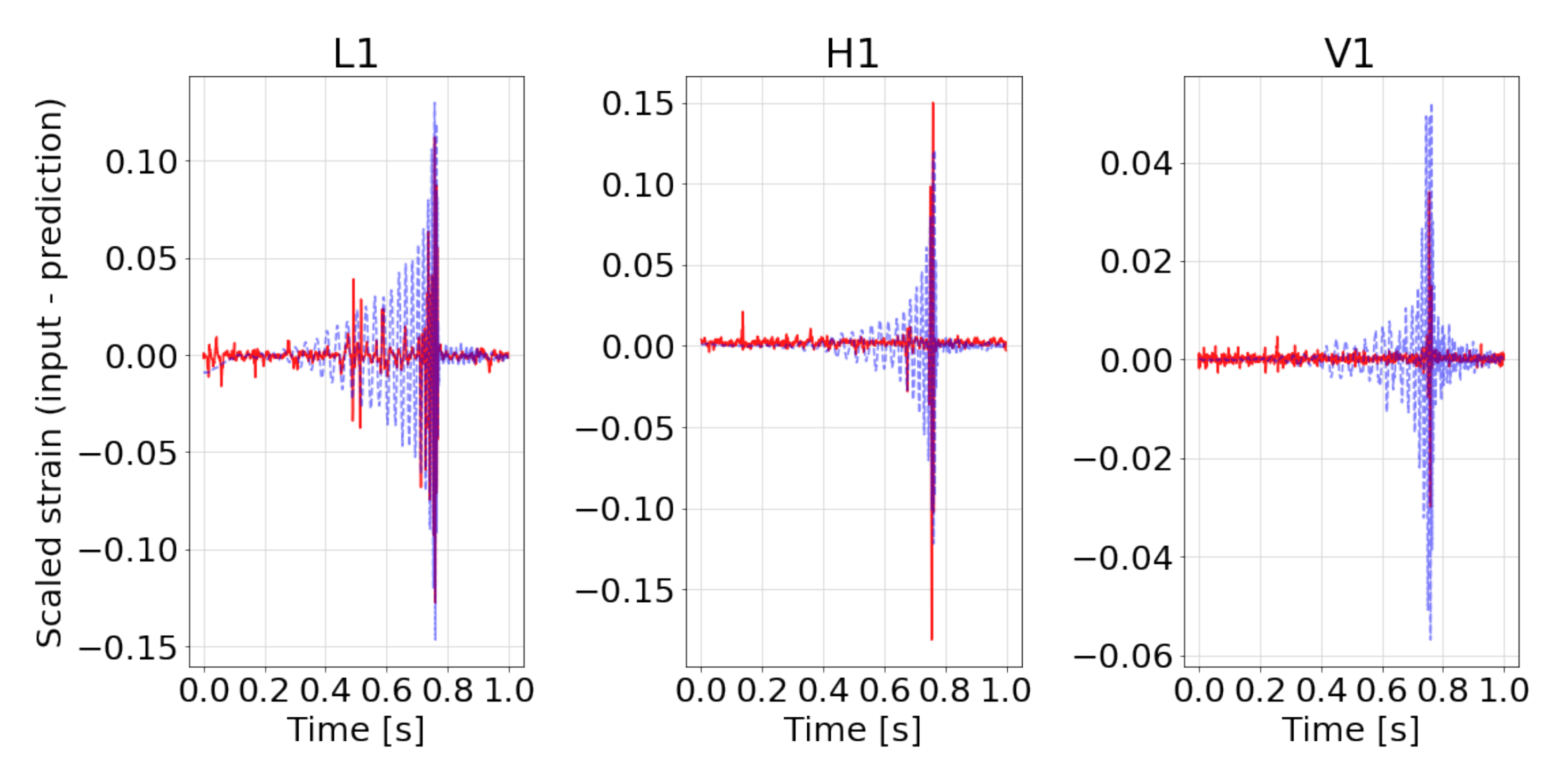}
  \caption{Three examples of GW reconstruction using AE. Presented plots were generated after subtracting from the AE predictions, input signal (red curve). For the comparison we added the expected signal using blue, dashed curve (difference between the input signal and the ground-truth output signal). From left to right: results for $L1$, $H1$ and $V1$ detectors.
}    
  \label{fig:real_sample_rec} 
\end{figure}

The ability of the AE to reconstruct the detectors' noise was investigated as previously in the case of simulated data. The differences between the input strain and AE reconstructions were compared with the expected values. Examples  of this comparison are presented in Figure \ref{fig:real_sample_rec}. The summary of the match between the injected and the reconstructed waveforms is presented in Appendix \ref{appendix:match}. The AE trained on $L1$ data achieved the best results, since the GW injected into the strain were extracted to the greatest extent in the merger part and partially in the inspiral part. On the other hand, the AE trained on the other datasets reconstructed mainly the merger part. Overall, the AE seemed to fail to reconstruct lower amplitudes and frequencies of the GW signal (the inspiral and ringdown part).

\begin{figure}[htbp]
  \includegraphics[width=\columnwidth]{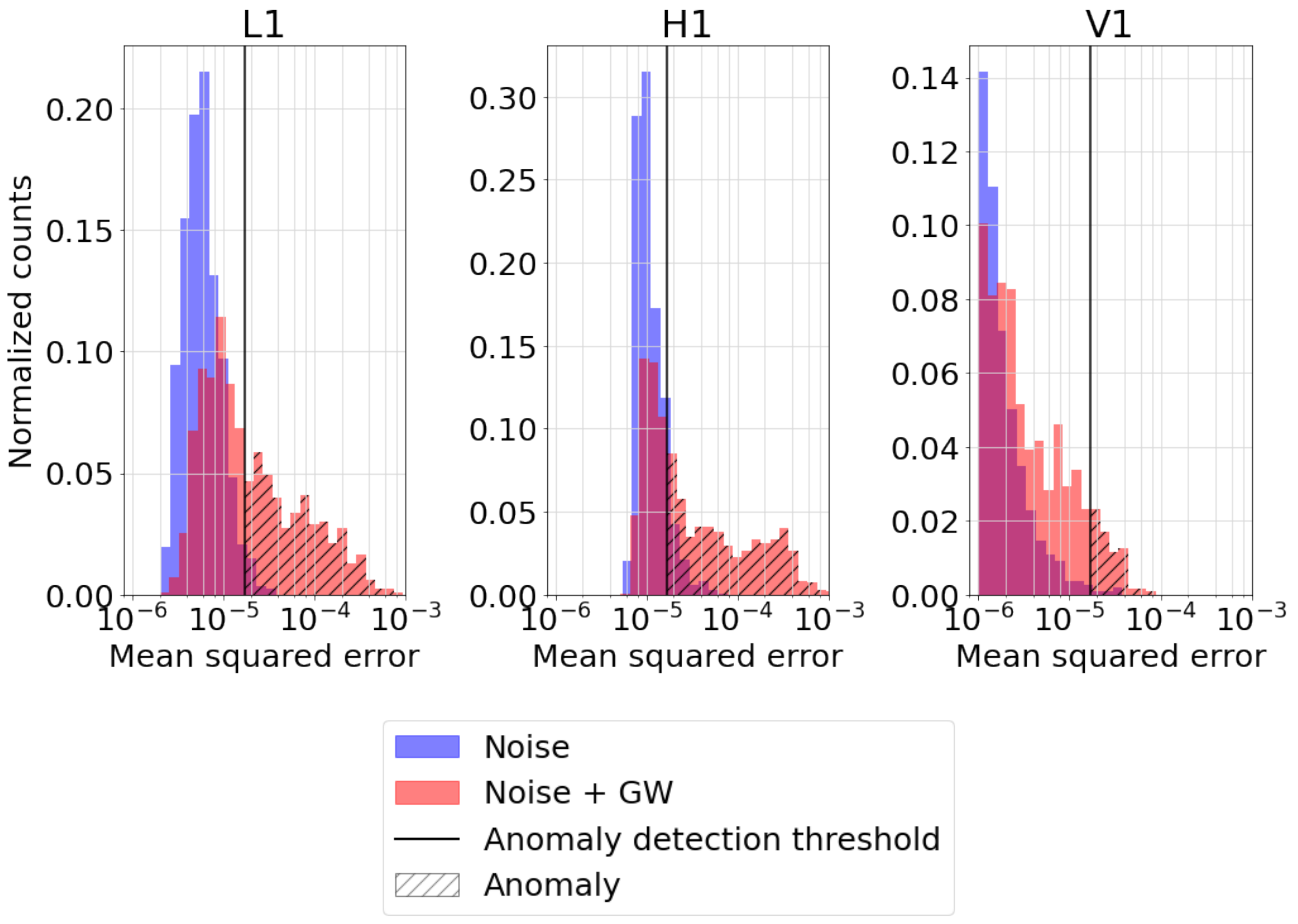}
  \caption{Distribution of MSE between the AE predictions and the input strain for two types of studied signals: injected GW (red histograms) and noise (blue histograms). Plots correspond to the $L1$ (\textit{left plot}), $H1$ (\textit{middle plot}) and $V1$ (\textit{right plot}) datasets. Additional black vertical line representing anomaly $DT$ was added to emphasize amount of detected anomalies (hatched areas).
}    
  \label{fig:real_hist} 
\end{figure}

To compute the threshold for the anomaly detection, we generated histograms of the MSE for every detectors' dataset and compared its values with the FPR. The results are shown in Figure \ref{fig:real_hist}. The anomalies covered a greater range of MSE values than the noise, with an overlapping range varying for different datasets (coloured in burgundy in Figure \ref{fig:real_hist}). In the case of $L1$ data, this region was the smallest whereas for $V1$, it was the largest. 
As in the case of simulated data, we defined a detection threshold for anomalies assuming FPR=5\%, which resulted in the following thresholds: $DT_{L1}=1.3\cdot10^{-5}$ for $L1$, $DT_{H1}=2.2\cdot10^{-5}$ for $H1$ and $DT_{V1}=4.3\cdot10^{-6}$ for $V1$.
The results of the anomaly searches on real datasets are present in Table \ref{tab:real_results} in the form of confusion matrix.
Anomaly row relates to the hatched area on panels in Figure \ref{fig:real_hist}. In the case of $L1$ and $H1$ datasets, around half of detected anomalies are correctly related to the injected GW. 
Samples that did not exceed corresponding $DT$ had low SNR. That was also the case of the $V1$ dataset where only 27 \% of samples exceeded $DT_{V1}$.  Details presenting the relation between the SNR of the injected GW and MSE of the reconstructed waveform for the real datasets can be found in the Appendix \ref{appendix:snr_mse}.

\begin{table*}[t]
  \centering 
  \begin{tabular}{| c || c | c || c | c || c | c |} 
    \hline
    \multirow{2}{*}{} & \multicolumn{2}{c||}{$L1$} & \multicolumn{2}{c||}{$H1$} & \multicolumn{2}{c|}{$V1$} \\ \cline{2-7}
      & Inj. GW & Noise
      & Inj. GW & Noise
      & Inj. GW & Noise
      \\ \hline\hline 
    Anomaly & 52 \% & 5 \% & 50 \% & 5 \% & 27 \% & 5 \%\\ \hline 
    Non-anomaly & 48 \% & 95 \% & 50 \% & 95 \% & 73 \% & 95 \% \\ \hline
  \end{tabular}
\vskip 1em  
\caption{Results of anomaly detection of CNN-AE at FPR=5\% for $L1$, $H1$ and $V1$ datasets in the form of confusion matrix. Columns relate to the ground-truth values whereas rows to the predictions. For all datasets significant majority of detected anomalies correctly corresponded to the data instances with injected GW. However, more than a third part of non-anomalous class (samples that did not exceed DT for a given detector) related to the low SNR injected GW (see Fig. \ref{fig:overview} for comparison between $L1$, $H1$ and $V1$ GW SNR distributions). }  
\label{tab:real_results}
\end{table*}

\begin{figure}[htbp]
  \includegraphics[width=0.5\columnwidth]{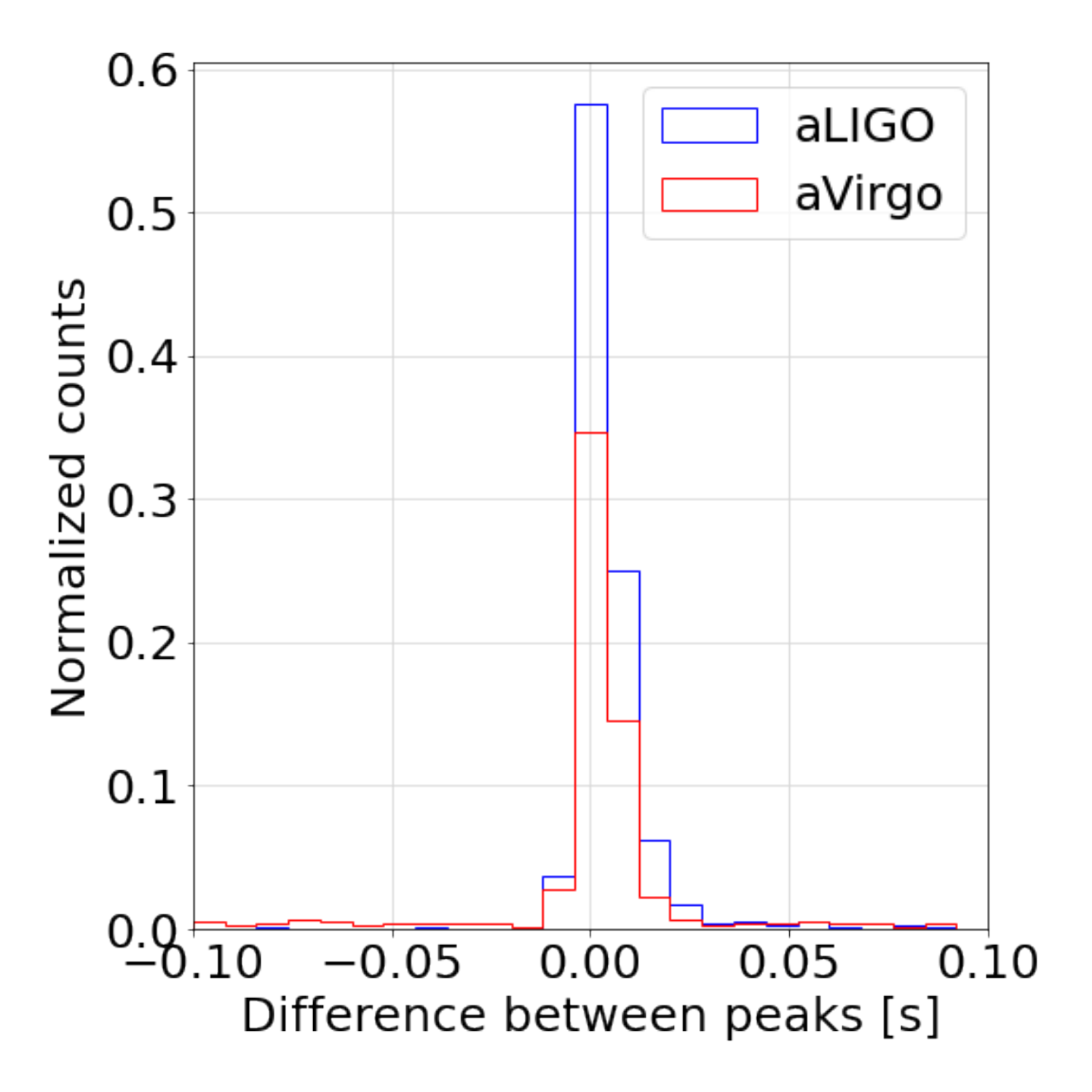}
  \hskip 10pt
  \includegraphics[width=0.5\columnwidth]{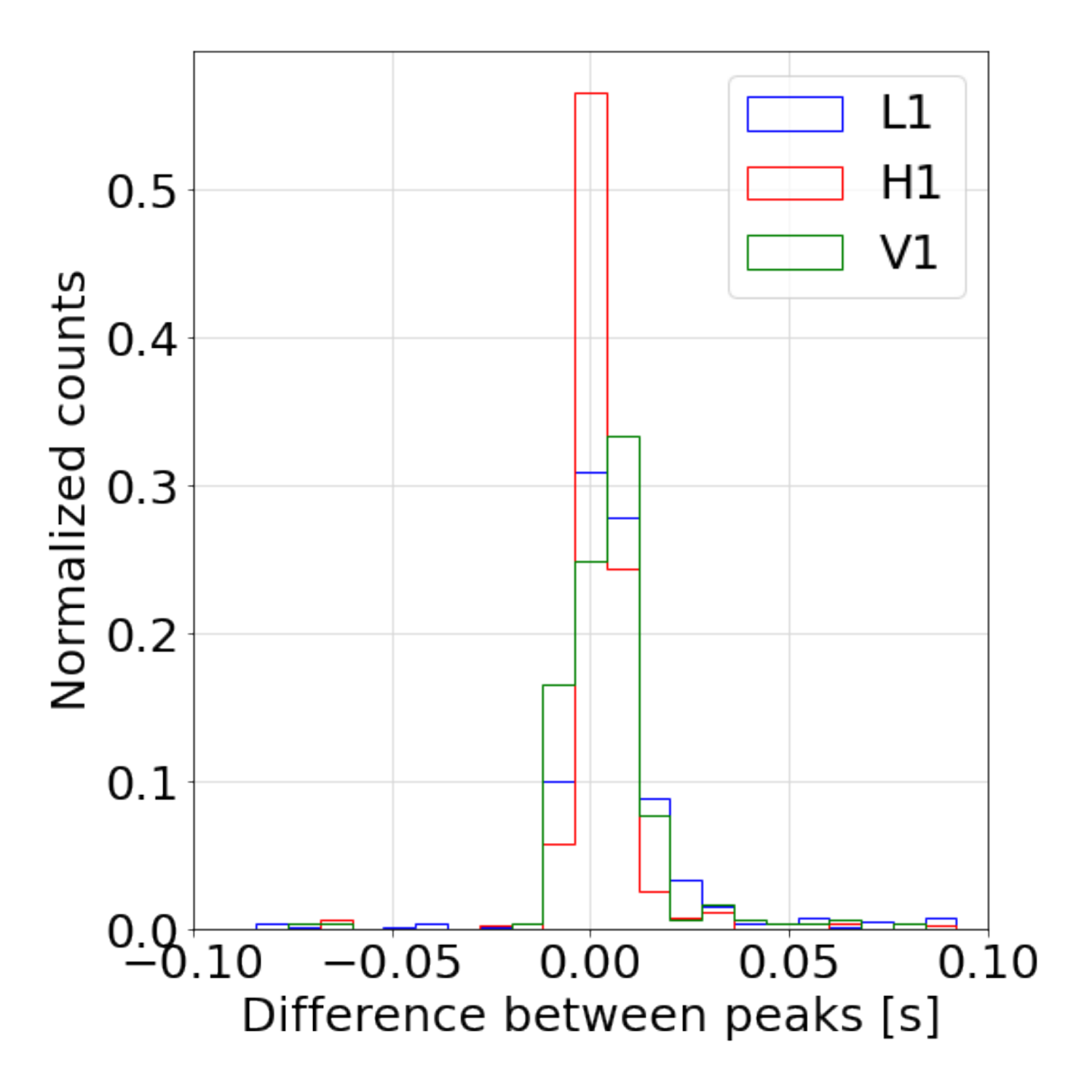}
  \caption{Anomaly localisation based on the difference in peak positions between reconstructed and injected signals. \textit{Left plot}: The results for the simulated datasets. \textit{Right plot}: The results for the real datasets. 
}    
  \label{fig:position} 
\end{figure} 

Additionally, we performed tests on the localisation in time of anomalies detected by our method. We compared the known times of the GW injection into the data with the time of the reconstructed signal. Specifically, we subtracted the times corresponding to the maximum amplitude peaks of both signals, and plotted histograms of the resulting differences for all the anomalies that exceeded computed previously DT. For comparison, we performed this procedure also for the simulated datasets. The results are shown in Figure \ref{fig:position}. In all the studied cases, around $95\%$ of detected anomalies were localised within $0.05$ s intervals around the injection times. We conclude that this feature may be useful not only for detection and reconstruction, but also potentially for other applications, such as the localisation of signals in the data from many detectors, and subsequent sky localisation of the sources.

\subsection{Anomaly searches on confirmed GW detections}
\label{ssec:res_gw}

Using a selection of real GW detections provided by the LIGO-Virgo collaboration on the GWOSC platform \cite{gwosc}, we tested the AE on three relatively strong signals from the GWTC-1 O1-O2 catalog \cite{2019PhRvX...9c1040A}: GW150914 \cite{GW150914}, GW170608 \cite{2017ApJ...851L..35A} and GW170814 \cite{PhysRevLett.119.141101}. The reported network SNR (square root of a sum of squares of SNRs from individual detectors $\rho_i$) $\rho_{net} = \sqrt{\sum_i \rho^2_i}$ is ${\simeq}24$ for GW150914, ${\simeq}15$ for GW170608 and ${\simeq}16$ for GW170814 \cite{gwosc}. Assuming two equally sensitive detectors, each of them would measure SNR of ${\simeq}17$ for GW150914, ${\simeq}10$ for GW170608 and and ${\simeq}11$ for GW170814; note that the GW170814 was a three-detector event, with the single-detector SNRs in $H1$, $L1$ and $V1$ equal to 7.3, 13.7, and 4.4, respectively \cite{PhysRevLett.119.141101}. In reality, due to differences in sensitivity the detectors registered the signals with different SNRs, which were nevertheless near the single-detector SNR detection threshold, established by the FPR=5\% condition.

\begin{figure}[htbp]
  \centering
  \includegraphics[width=0.8\columnwidth]{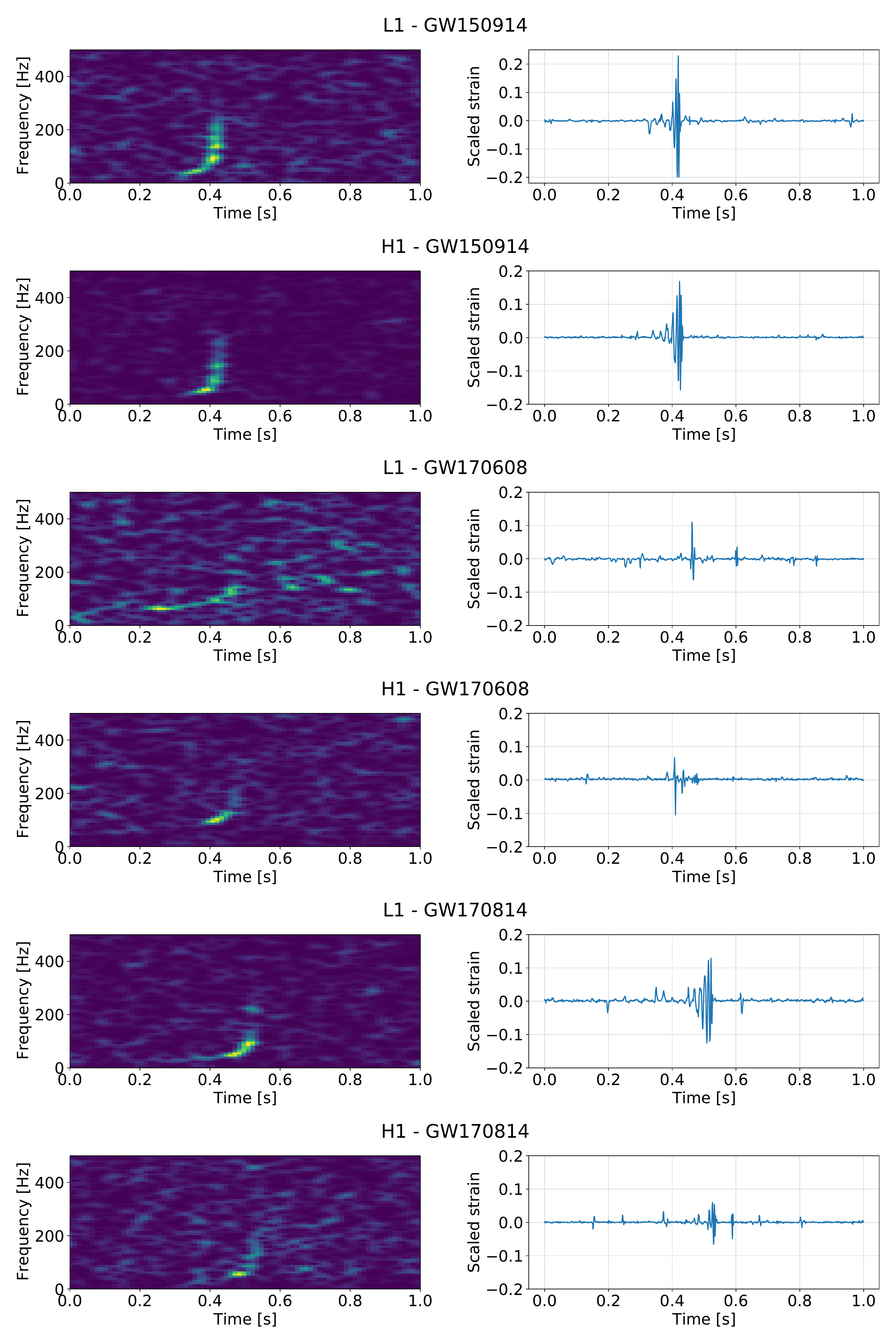}
  \caption{Test of the CNN-AE on the dataset containing confirmed detections using $L1$ and $H1$ datasets for GW150914 (four upper plots), GW170608 (four middle plots) and GW170814 (four bottom plots). \textit{Left plots}: spectrograms presenting the relation between frequency and time for the segment of real data containing GW. \textit{Right plots}: the difference between the CNN-AE predictions and the input data. 
}    
  \label{fig:conf_samples} 
\end{figure} 

After whitening, test data were fed into the AE. Then the reconstructed values were subtracted from the input data. The results of subtraction for GW150914 are shown on the two upper, right plots in Figure ~\ref{fig:conf_samples} for the both LIGO detectors. The presented signals resulted in the $MSE_{L1}=3.2\cdot10^{-4}$ and  $MSE_{H1}=1.0\cdot10^{-3}$. For both detectors, MSE had significantly higher values than the corresponding detection thresholds of FPR=5\%.

In case of GW170608, AE detected the event; however, the reconstructed signal was limited to the merger part for both the LIGO detectors as shown on the two middle, right plots in Figure ~\ref{fig:conf_samples}. A potential explanation for this weaker reconstruction of this particular GW is related to the different mass ranges of GW170608 and the GWs used for the training of the AE. GW170608 had BBH component masses corresponding to around $m_1=11.0$ and $m_2=7.6$ $M_\odot$ which were substantially smaller than the masses used in the data generation (see Section \ref{sec:meth} for more details). We note that the H1 detector was nominally outside of observing mode at the time of this event. This data release includes H1 data around the time of this event, by using a modified segment list, as was done for the published analysis \cite{2017ApJ...851L..35A,gwtc1,gwosc}. 
Nevertheless, the AE detected GW170608, proving its generalisation capabilities towards recognising gravitational waveforms it was not trained for. This is a sure advantage of our proposed method, when compared to the matched filtering method. Furthermore, the MSE values for both detectors had slightly higher values above the detection thresholds at FPR=5\%: $MSE_{L1}=5.3\cdot10^{-5}$ and  $MSE_{H1}=4.1\cdot10^{-5}$.

The last studied test case, GW170814, was detected in both LIGO detectors with a substantial part of the waveform being recovered, as shown on the two bottom, right plots in Figure \ref{fig:conf_samples}. For the $V1$ data, AE failed to detect the event evidently due to low reported SNR of ${\simeq}4$ \cite{PhysRevLett.119.141101}. The MSE for the $L1$ and $H1$ datasets were above the data threshold, at $5\%$ and equal to  $MSE_{L1}=2.2\cdot10^{-4}$, $MSE_{H1}=2.2\cdot10^{-4}$, whereas for $V1$ the MSE was below the threshold: $MSE_{V1}=1.8\cdot10^{-6}$.

These results confirm that our proposed method of using CNN-AE for anomaly searches is able to detect real GWs, even though the deep learning model was trained on relatively uncomplicated datasets, based on the information related to a particular GW waveform models, varying with respect to the limited range of masses and distances.

\section{Summary}
\label{sec:summary}

In this paper, we proved that AEs are a potentially powerful method for anomaly searches in the GW data. A relatively simple AE, consisting of only three hidden layers, was capable of detecting anomalies, defined in terms of transient BBH GWs (as well glitches in real data), and trained on either simulated or real data. Moreover, our proposed method was able to detect as anomalies all three confirmed GWs used as a test case and even partially reconstruct the waveforms of the underlying signals.

In the proposed method we introduced a metric allowing for the automatic detection of anomalies. The chosen metric was MSE. Using this metric, we defined the threshold for the anomaly detection by associating MSE with FPR. Presented in the manuscript results referred to FPR=5\%. At such a threshold we were able to detect in case of LIGO data, almost all of the injected GW into the simulated dataset, as well as around 50\% for real detector data. In the case of Virgo, for simulated data around half of injected signals were detected and quarter for real data. The reason for worse results on real data taken from O1-O2 observational run, was substantial difference in the sensitivity with respect to the simulated counterpart. The sensitivity of real detectors were overall worse than the designed one. However, the real sensitivity significantly improved for O3 observational run after detector upgrades. Once the data from O3 becomes public, the anomaly searches of proposed CNN-AE method are expected to improve.

Our method also proved to be useful in the precise time localisation of anomalies. Anomalies exceeding the detection threshold for all the studied datasets (real and simulated) turned out to be localised with the accuracy below 0.05 second. Localisation of the peak amplitude may be particularly useful for the localisation of GW sources on the sky in the multi-detector analysis. However, such an application of our method requires a separate study.

Finally, the successful detection of GW170608 proved the generalisation capabilities of our AE, towards the detection of GWs with parameters that are different than those of the GWs used for the AE training and in the data nominally outside of the observing mode. 

Among projects we are considering are applications of recurrent neural networks instead of CNN in the anomaly searches as an alternative suited for the time-series data as well as anomaly searches of different GW types such as core collapse supernova signals.

\ack
The work was partially supported by the Polish NCN grants no. 2016/22/E/ST9/00037, 2017/26/M/ST9/00978 and 2020/37/N/ST9/02151, as well as the European Cooperation in Science and Technology COST action G2Net (CA17137). The Quadro P6000 used in this research was donated by the NVIDIA Corporation. The production computations were supported in part by PL-Grid Infrastructure and by the MNiSW grant for expansion of the strategic IT infrastructure for science.

This research has made use of data, software and/or web tools obtained from the Gravitational Wave Open Science Center (https://www.gw-openscience.org/ ), a service of LIGO Laboratory, the LIGO Scientific Collaboration and the Virgo Collaboration. LIGO Laboratory and Advanced LIGO are funded by the United States National Science Foundation (NSF) as well as the Science and Technology Facilities Council (STFC) of the United Kingdom, the Max-Planck-Society (MPS), and the State of Niedersachsen/Germany for support of the construction of Advanced LIGO and construction and operation of the GEO600 detector. Additional support for Advanced LIGO was provided by the Australian Research Council. Virgo is funded, through the European Gravitational Observatory (EGO), by the French Centre National de Recherche Scientifique (CNRS), the Italian Istituto Nazionale di Fisica Nucleare (INFN) and the Dutch Nikhef, with contributions by institutions from Belgium, Germany, Greece, Hungary, Ireland, Japan, Monaco, Poland, Portugal, Spain.

\section*{Data Availability Statement}
The data that support the findings of this study are available from the corresponding author upon reasonable request.

\section*{References}
\bibliography{bibfile}

\clearpage

\begin{appendices}

\section{Match/overlap between the injected and reconstructed waveforms}
\label{appendix:match}

\subsection{Simulated dataset}

\begin{figure}[htbp]
 \includegraphics[width=\columnwidth]{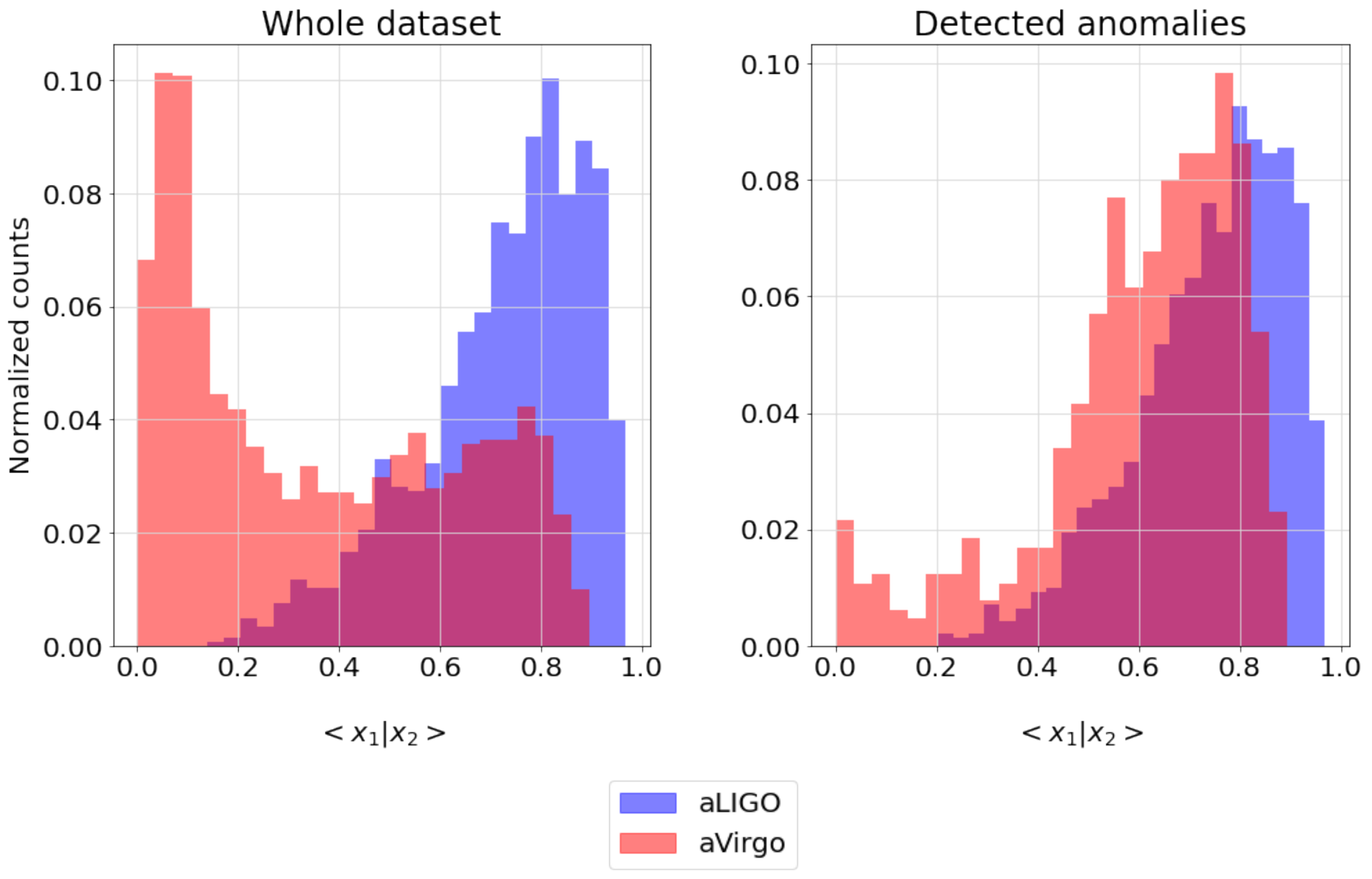}
 \caption{Distribution of a match between the injected ($x_1$) and the reconstructed ($x_2$) waveforms as a function of normalized scalar product $<x_1 | x_2>$. Scalar product equals to one relates to the full match between waveforms whereas value of zero - no match. \textit{Left plot}: results for the whole studied dataset of particular detector; \textit{right plot}: results of a match after applying the anomaly detection threshold at FPR=5\%.
  }    
  \label{fig:sim_match} 
\end{figure}

To measure the match between the injected and the reconstructed waveforms we used the normalized scalar product $<x_1 | x_2>$ in the time domain resulting in values in a range $(0, 1)$. Zero related to no match, whereas one to full match between the waveforms. Presented results in Fig. \ref{fig:sim_match} corresponds to the whole studied dataset for aLIGO and aVirgo detectors (left panel) as well as samples exceeding the anomaly detection thresholds (right panel). Applying respective $DT$ allowed to substantially reduce number of samples with no match between the waveforms as a result of low SNR of injected GW. Samples exceeding $DT$ were reconstructed to a greater extent which resulted in $<x_1 | x_2>$ closer to one.

\subsection{Real dataset}

\begin{figure}[htbp]
 \includegraphics[width=\columnwidth]{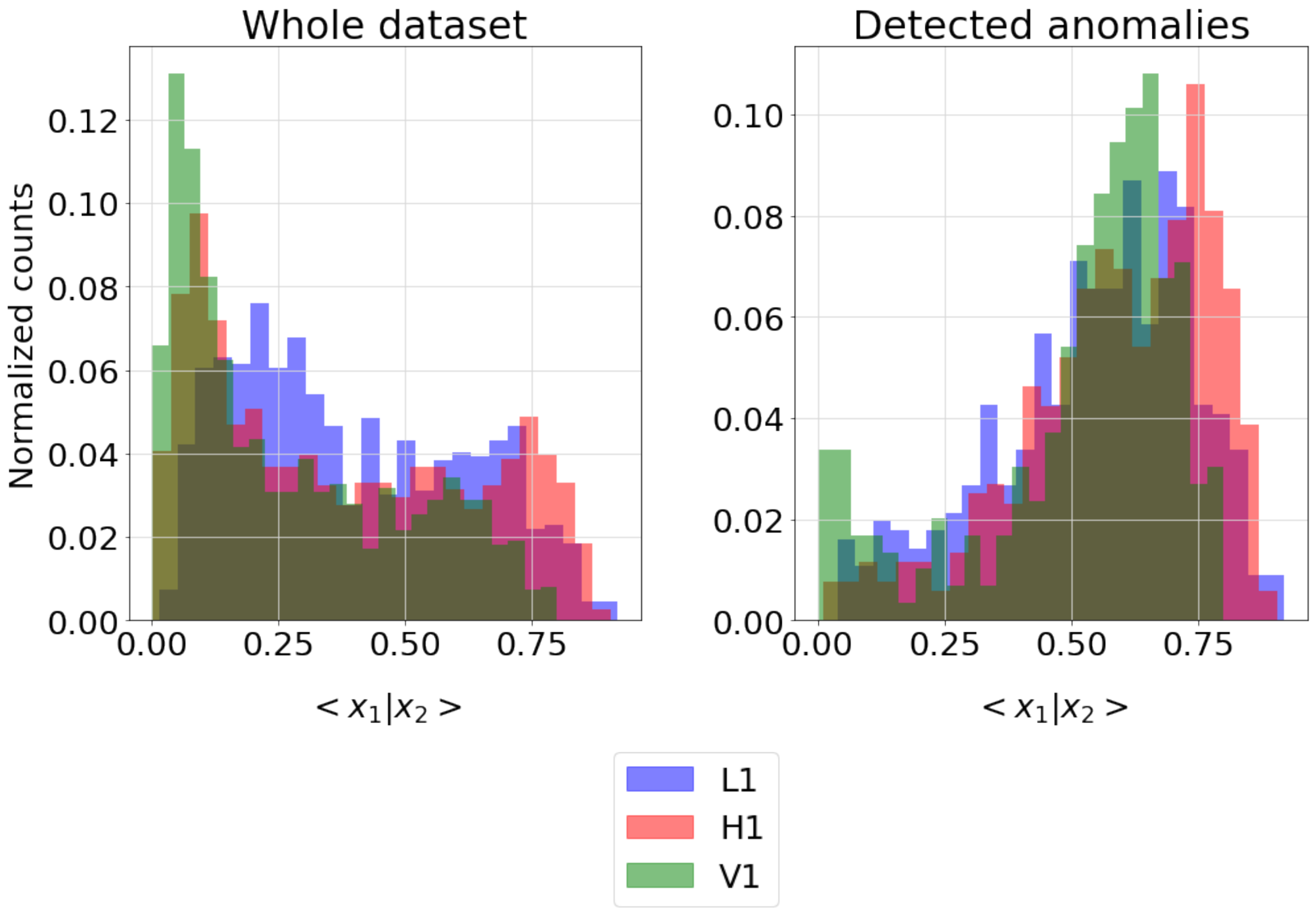}
 \caption{Distribution of a match between the injected ($x_1$) and the reconstructed ($x_2$) waveforms as a function of normalized scalar product $<x_1 | x_2>$. Scalar product equals to one relates to the full match between waveforms whereas value of 0 - no match. \textit{Left plot}: results for the whole studied dataset of particular detector; \textit{right plot}: results of a match after applying the anomaly detection threshold at FPR=5\%.
  }    
  \label{fig:real_match} 
\end{figure}

The same metric as in case of simulated dataset was used to study the match between the injected and the reconstructed waveforms for the real datasets. Presented in Fig. \ref{fig:real_match} results show the similarities in the match between consecutive datasets ($L1$, $H1$ and $V1$) as well as the effect of applying the anomaly detection threshold. Samples with $<x_1 | x_2>$ close to zero had low SNR. As a result they were poorly reconstructed which in turn translated into low value of MSE. Samples exceeding respective $DT$ were reconstructed to a greater extent as in the case of simulated data. However, overall match was worse - the mean values of $<x_1 | x_2>$ for real datasets were around 0.6, whereas for simulated data 0.8.

\clearpage
\section{Signal-to-Noise Ratio vs Mean Squared Error}
\label{appendix:snr_mse}

\subsection{Simulated dataset}

For the aLIGO dataset above SNR=20, the MSE-SNR relation was almost linear, with a small spread of individual data instances along MSE. Whereas with the decline of SNR, the spread of MSE significantly increased, characterized by the non-linearity in the MSE-SNR relation. Anomalies around the same SNR for values below $10$ varied up to an order of magnitude in MSE. Manual inspection of data samples containing anomalies of low SNR provided an explanation of this behaviour. In the analysed samples, only the merger part of the gravitational waveform was recovered. For lower SNRs (below $10$) the recovery was partial and dependent on the variability of the noise. If the amplitude of the noise in a given data segment was small comparing to the injected GW signal, the resulting MSE was higher than in the case of noise samples with larger amplitudes. Overall, for the aLIGO dataset, the susceptibility of AE to the local variability of the noise was inversely proportional to the SNR of the injected anomaly.

In case of the aVirgo dataset, the mentioned susceptibility was more significant. Matched filter SNRs for all injected GWs covered a smaller range of values than for aLIGO (compare SNR ranges on the bottom right plot in Figure \ref{fig:overview} for aLIGO and aVirgo datasets). In the majority of studied cases, the recovery of the anomaly was partial and limited to the merger part. 

\begin{figure}[htbp]
 \includegraphics[width=\columnwidth]{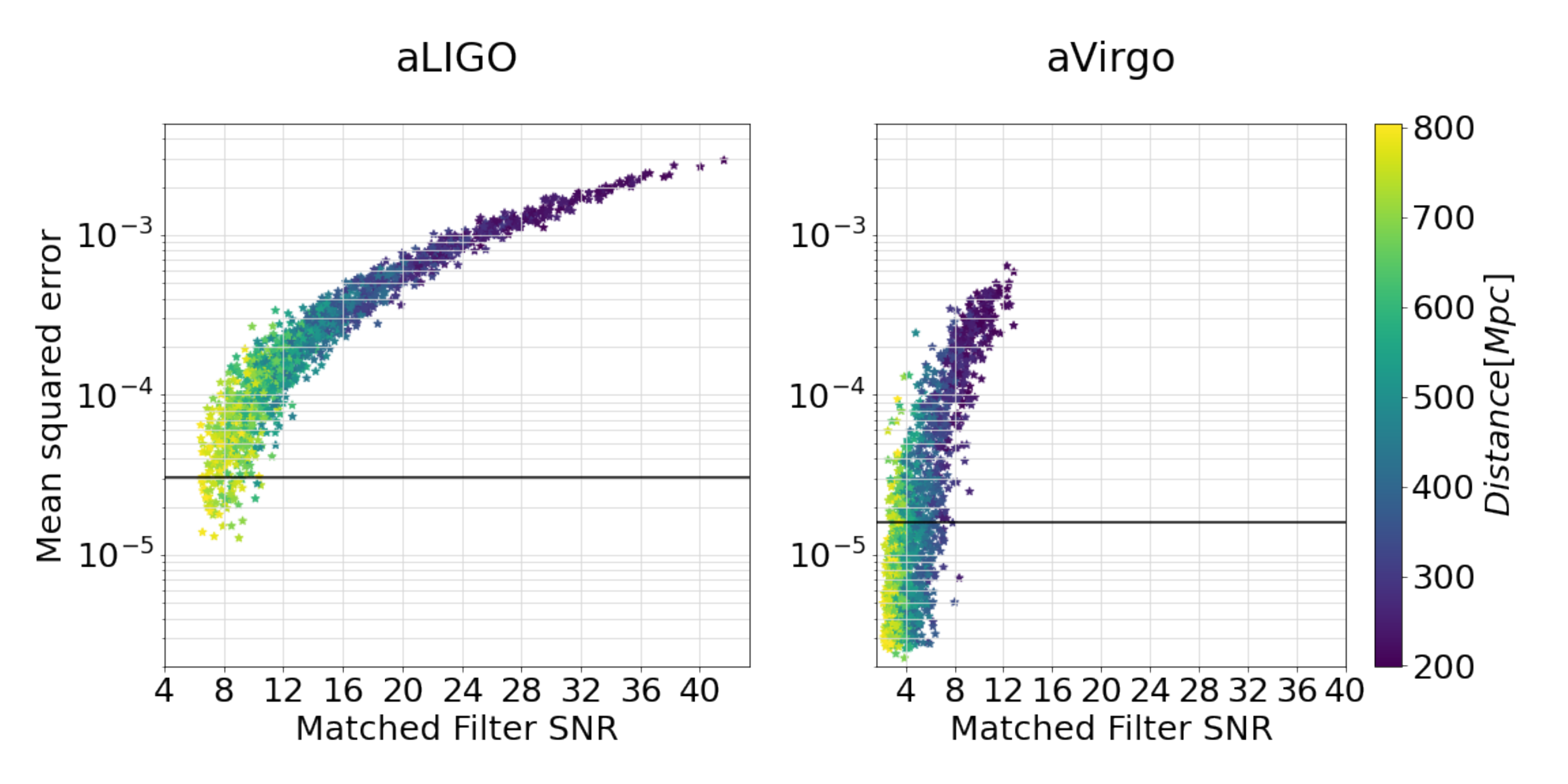}
  \caption{Relation between MSE and matched filter SNR for aVirgo and aLIGO datasets colored with a distance to the GW source. Black horizontal line corresponds to the detection threshold of anomalies at FPR=5\%.
  }    
  \label{fig:sim_ad} 
\end{figure} 

\subsection{Real dataset}

The relation between SNR and MSE presented similar features as for the simulated data discussed above. The variability of MSE for samples of similar SNR was largest among the weakest anomalies. The increase of the matched filter SNR led to the decrease of the spread in MSE, and thus their relation became more linear. This was the case for the AEs trained on the $L1$ and $H1$ datasets. In contrast, the AE trained on the $V1$ dataset was more susceptible to the local variability of the noise. Since the injected anomalies had a small SNR for $V1$ (majority of injected GW had SNR below $10$), their amplitude was significantly lower than the detectors noise. As a result, the anomaly detection depended on the variability of the noise, which had a random character. 

\begin{figure}[htbp]
  \includegraphics[width=\columnwidth]{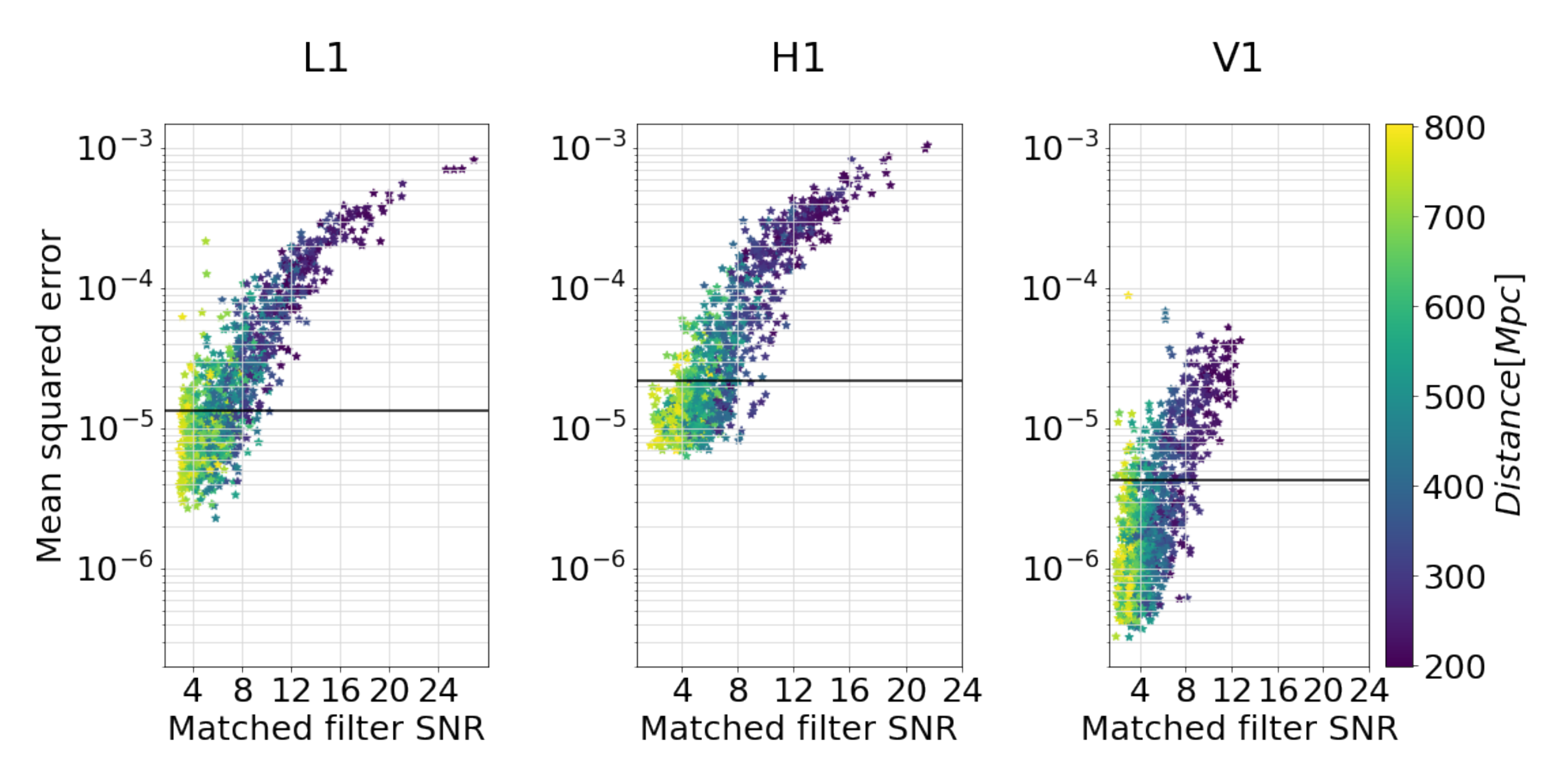}
  \caption{Relation between the MSE and the matched filter SNR for $L1$ (\textit{left plot}), $H1$ (\textit{middle plot}) and $V1$ (\textit{right plot}) datasets colored with a distance to the GW source. Black horizontal line corresponds to the detection threshold of anomalies at FPR=5\%.
}    
  \label{fig:real_ad} 
\end{figure}

\end{appendices}

\end{document}